\documentclass[notitlepage,prd,showkeys,preprintnumbers,floatfix,superscriptaddress,nofootinbib]{revtex4-1}

\usepackage{float}
\usepackage{amsmath}
 
\usepackage{amssymb}
\usepackage{graphicx}
\usepackage{mathtools,slashed}
\usepackage{bm}
\usepackage{comment}
\usepackage{xcolor}
\usepackage{subfigure}
\usepackage{array}
\usepackage{multirow}
\usepackage{url}
\usepackage{color}
\usepackage[T1]{fontenc}
\usepackage{graphicx}
\usepackage{esint}
\usepackage{hyperref}
\usepackage{atbegshi}
\usepackage{lipsum}
\usepackage{braket}

\begin{document}

 \preprint{\vbox{
\hbox{JLAB-THY-24-4077} 
}}

\title{
Inclusive reactions from finite Minkowski spacetime 
correlation functions
}

\author{Marco A. Carrillo}
\email[]{mcarr020@odu.edu}
\affiliation{Department of Physics, Old Dominion University, Norfolk, Virginia 23529, USA}
\affiliation{Thomas Jefferson National Accelerator Facility, 12000 Jefferson Avenue, Newport News, Virginia 23606, USA}

\author{Ra\'ul A.~Brice\~no}
\email[]{rbriceno@berkeley.edu}
\affiliation{Department of Physics, University of California, Berkeley, CA 94720, USA}
\affiliation{Nuclear Science Division, Lawrence Berkeley National Laboratory, Berkeley, CA 94720, USA}

\author{Alexandru M. Sturzu}
\email[]{amsturzu@wm.edu}
\affiliation{Department of Physics, College of William \& Mary, Williamsburg, VA 23187, USA }

\begin{abstract}

The need to determine scattering amplitudes of few-hadron systems for arbitrary kinematics expands a broad set of subfields of modern-day nuclear and hadronic physics. In this work, we expand upon previous explorations on the use of real-time methods, like quantum computing or tensor networks, to determine few-body scattering amplitudes. Such calculations must be performed in a finite Minkowski spacetime, where scattering amplitudes are not well defined. Our previous work presented a conjecture of a systematically improvable estimator for scattering amplitudes constructed from finite-volume correlation functions. Here we provide further evidence that the prescription works for larger kinematic regions than previously explored as well as a broader class of scattering amplitudes. Finally, we devise a new method for estimating the order of magnitude of the error associated with finite time separations needed for such calculations. In units of the lightest mass of the theory, we find that to constrain amplitudes using real-time methods within $\mathcal{O}(10\%)$, the spacetime volumes must satisfy $mL \sim \mathcal{O}(10-10^2)$ and $ mT\sim \mathcal{O}(10^2-10^4)$.

\end{abstract}

\keywords{Lattice QCD, finite Minkowski spacetime, quantum computing}

\nopagebreak

\maketitle

\section{Introduction}

Whether one is interested in hadron spectroscopy~\cite{Lebed:2022vfu, Achenbach:2023pba, Bulava:2022ovd}, hadron and nuclear structure~\cite{Drischler:2019xuo, Brambilla:2022ura}, or precision electroweak processes and beyond the Standard Model (BSM) searches~\cite{DellOro:2016tmg, Dolinski:2019nrj, Aoyama:2020ynm, Acharya:2023swl}, there is a broad set of motivations for studying scattering amplitudes involving few-hadron systems. Generally speaking, in each one of these fields most observables can be reconstructed from two broad classes of amplitudes, purely hadronic scattering amplitudes and amplitudes involving initial and final hadronic states that are coupled via electroweak and/or BSM probes that can be introduced perturbatively. In some fields, e.g. hadron spectroscopy, the goal is to constrain the hadronic amplitude to search for a signal of excited states, which appear as unstable resonances in scattering amplitudes. Another example is that of hadronic weak decays, where the goal is to constrain quantities such as the Cabibbo–Kobayashi–Maskawa matrix elements \cite{FLAG:2021npn}, which are generally clouded by dominant contributions from quantum chromodynamics (QCD).

Given the non-perturbative nature of strong interactions, any systematic method for constraining few-hadron reactions must treat the theory exactly. To this day, Lattice QCD is the only non-perturbative tool that has provided constraints on hadronic scattering amplitudes directly from QCD~\cite{Briceno:2017max}. Recent examples of the classes of scattering amplitudes being determined via lattice QCD include purely hadronic resonant multichannel~\cite{Dudek:2014qha,Wilson:2014cna,Briceno:2017qmb,Woss:2019hse}, involving electromagnetic currents~\cite{Briceno:2016kkp,Radhakrishnan:2022ubg,Boyle:2022uba}, and three-particle scattering amplitudes~\cite{Hansen:2020otl}. 

Since Lattice QCD places the theory in a finite discretized Euclidean spacetime, scattering amplitudes are na\"ively inaccessible for two reasons. First, the Euclidean nature of the calculations generally prohibits the usage of the Lehmann-Symanzik-Zimmermann (LSZ) reduction formula to obtain scattering amplitudes directly from correlation functions. Secondly, asymptotic states, as required in the definition of scattering states, can not be rigorously defined in a finite volume. 

The progress made in the field is in large part due to the extensive and growing literature~\cite{Luscher:1986pf,Luscher:1990ux,Rummukainen:1995vs,Bedaque:2004kc,Feng:2004ua,Kim:2005gf,He:2005ey,Lage:2009zv,Davoudi:2011md,Fu:2011xz,Leskovec:2012gb,Briceno:2012yi,Li:2012bi,Briceno:2013lba,Guo:2012hv,Briceno:2014oea,Christ:2005gi, Polejaeva:2012ut,Briceno:2012rv,Hansen:2014eka,Hansen:2015zga,Hammer:2017uqm,Hammer:2017kms,Guo:2017ism,Mai:2017bge,Briceno:2017tce,Doring:2018xxx,Briceno:2018mlh,Briceno:2018aml,Blanton:2019igq,Jackura:2022gib,Dawid:2023jrj,Dawid:2023kxu, Christ:2005gi,Lellouch:2000pv,Meyer:2011um,Bernard:2012bi,Hansen:2012tf,Agadjanov:2014kha,Briceno:2014uqa,Feng:2014gba,Briceno:2015csa,Briceno:2015tza,Baroni:2018iau,Briceno:2019opb, Raposo:2023oru} dedicated to deriving the relation between finite- and infinite-volume observables. The most well-known example is the relationship between two-particle finite-volume energy levels and infinite-volume purely hadronic scattering amplitudes \cite{Luscher:1986pf, Luscher:1990ux}. Similarly, one can construct relations between finite- and infinite-volume matrix elements of external currents \cite{Lellouch:2000pv}. To use the aforementioned finite-volume formalism in the analysis of lattice QCD correlation functions, it is also necessary to develop a parallel formalism that describes the analytic structure of infinite-volume amplitudes, which has motivated various efforts~\cite{Briceno:2020vgp,Sherman:2022tco, Jackura:2020bsk,Dawid:2023jrj,Jackura:2023qtp}. 

Overall, these formalisms require isolating all power-law finite-volume effects and physical singularities of finite- and infinite-volume correlators, respectively. Both of these effects are due to intermediate particles going on their mass shell. As we consider increasingly higher energies this methodology becomes impractical due to two primary reasons. First, as the number of particles that can go on-shell increases, the derivation of these relations becomes more challenging. Second, as the energy increases, the number of singularities and, in general, the number of degrees of freedom increase. These two issues combined make the analysis of any scattering observable increasingly harder and perhaps unsystematic for higher energies. 

Given the aforementioned challenges, here we explore an alternative approach for obtaining scattering amplitudes non-perturbatively that do not require the finite-volume formalisms mentioned above. In particular, we consider a general and simple method for estimating infinite-volume scattering observables from finite Minkowski spacetime correlation functions, first introduced in Ref.~\cite{Briceno:2020rar}. This was inspired by exciting progress on novel computing frameworks, such as quantum algorithms ~\cite{Jordan:2012xnu, Marshall:2015mna, Klco:2018zqz, DeJong:2020riy, deJong:2021wsd, Bauer:2021gek, Nguyen:2021hyk, Andrade:2021pil, Davoudi:2021ney, Davoudi:2021ney, Araz:2022tbd, Jha:2023ecu, Jha:2023ump, Briceno:2023xcm, Lu:2019, Jha:2023, Klco:2020, Farrell:2023I, Farrell:2023II, farrell:2023,Farrell:2023fgd,Farrell:2024fit, Atas:2023, Klco:2019, Ciavarella:2023, RoggeroBaroni:2020, Roggero:2020, Baroni:2022, Sobczyk:2022, Hartse:2023, Bedaque:2022, Ciavarella:2020, Davoudi:2024wyv} and tensor networks ~\cite{Tang:2021uge}, for studying quantum field theories in a finite Minkowski spacetime.  

The estimators presented in Ref.~\cite{Briceno:2020rar} can be built using minimal assumptions and could be used to construct any desired amplitudes. In Sec.~\ref{sec:review}, we give a detailed definition of these estimators. Generally speaking, the definition of the estimators stems from the fact that any scattering amplitude can be constructed from Fourier transforms of time-dependent matrix elements. As is reviewed in detail in Sec.~\ref{sec:review}, in a finite volume the resultant functions have the wrong analytic properties. However, by shifting the arguments of the Fourier transform into the complex plane and performing simple averaging over different kinematic points, the resultant functions systematically reproduce the desired amplitudes using volumes that are one to two orders of magnitudes smaller than na\"ively needed. 

The main alternative to studying scattering processes via real-time methods, which requires the formation of wavepackets that are evolved in time, was proposed by Ref.~\cite{Jordan:2012xnu}. A major appeal for exploring the ``\emph{wavepacket approach}" is that it allows for an elegant visualization of the evolution of scattering states. At this stage, it is conceptually clear how this approach may be used for studying scattering processes involving two initial wave packets. This means that it is at least clear how to study purely hadronic scattering processes involving two-particle states in the initial and final states. Of course, one or both of these particles could be a bound state, e.g. a nucleus. In other words, the wavepacket approach can give access to QCD analogs of experimentally-accessible reactions. However, it is unclear how this procedure would be used to study two other classes of processes. The first are reactions where the electroweak or BSM sector can be introduced perturbatively. The second are experimentally accessible reactions involving three or more particles in the initial and final state, which are already being studied via lattice QCD~\cite{Hansen:2020otl}. The major advantage of the estimators over the wavepacket approach is that they give an exact prescription to obtain any $S$-matrix element of the desired theory, including reactions where external probes can be inserted perturbatively. 

Given that most of the progress in real-time methods has been made in lower-dimensional theories, Ref.~\cite{Briceno:2020rar} focused on the implication of estimators for amplitudes in 1+1D. Accordingly, in this work, we also focus our attention on strongly interacting systems in 1+1D. As a result, the remainder of the discussion in this and other sections will be restricted to 1+1D, even when it is not explicitly stated. However, it is worth emphasizing that the definitions of the estimators do not depend on the dimensions of the spacetime. While their predictive power can depend on the dimensionality, in this exploratory study, we do not consider their implications for higher dimensions.

Reference~\cite{Briceno:2020rar} paid close attention to a class of amplitudes that can be generally described as ``\textit{Compton-like}'' amplitudes. In this type of process, a single-particle state ($\varphi$) and a current ($\mathcal{J}$) scatter elastically, $\varphi +\mathcal{J}\to \varphi +\mathcal{J}$, where the current can serve as a perturbative proxy for an external field\footnote{If the current were the electromagnetic current, the resultant amplitude would be the Compton scattering amplitude for that scalar particle. We refer to this as a Compton-like amplitude because the process is similar to the physical Compton process but the current is a scalar.}. The aforementioned reference showed analytically and empirically that the finite-volume estimator effectively recovers the infinite-volume amplitude in a restricted kinematic region, where a single channel composed of two particles is kinematically open. 

In this work, we provide further empirical evidence of the effectiveness of the proposed estimator in two major ways. First, Ref.~\cite{Briceno:2020rar} explained that purely hadronic two-particle amplitudes could be obtained from {Compton-like} amplitudes using appropriately chosen currents and the LSZ formalism. In Sec.~\ref{sec:InfT}, we provide evidence that this is indeed possible for obtaining not just purely hadronic amplitudes but also transition amplitudes. Second, we extend the work done in Ref.~\cite{Briceno:2020rar} by demonstrating that this technique works when there are multiple channels kinematically open. This is shown for a few classes of examples in the same aforementioned section.

In particular, we demonstrate that one can directly constrain elements of scattering amplitudes in the presence of any number of kinematically opened channels. This is a major advantage over standard  methods constraining multi-channel amplitudes from finite-volume observables~\cite{Briceno:2012yi,Hansen:2012tf,Briceno:2014oea, Briceno:2015csa,Briceno:2021xlc}, where there is no one-to-one correspondence between finite- and infinite-volume quantities. In practice, one must resort to using parameterizations of scattering amplitudes and perform a global analysis of quantities constrained via lattice QCD~\cite{Wilson:2023anv,Wilson:2023hzu,Woss:2020ayi,Woss:2019hse,Woss:2018irj,Moir:2016srx,Dudek:2016cru,Wilson:2015dqa,Wilson:2014cna,Dudek:2014qha,Briceno:2017qmb,BaryonScatteringBaSc:2023ori,BaryonScatteringBaSc:2023zvt}, which results in a systematic error associated with the parametrizations used that is generally hard to assess. 

In addition to giving further evidence to the conjecture presented in Ref.~\cite{Briceno:2020rar}, here we raise an issue that has not been previously discussed. Although time is no longer imaginary, the Fourier transform discussed above na\"ively requires the integration to be performed over all time, but any realistic calculation can only be done for a finite number of time separations. In Sec.~\ref{sec:FiniteT}, we use the spectral decomposition of correlation functions to explore systematic errors associated with using a finite extent in time. For the sake of simplicity, we assume that the time evolution is continuous, although in general it will not be the case. Using these assumptions we find that for strongly interacting systems physical volumes of orders $mL \sim \mathcal{O}(10-10^2)$ and $ mT\sim \mathcal{O}(10^2-10^4)$ suffice to recover the infinite-volume amplitudes within reasonable error, with $m$ being the lightest mass of the theory.

\section{Review of formalism}\label{sec:review}

As previously discussed, in this work we show further evidence that the estimators proposed in Ref.~\cite{Briceno:2020rar} provide a systematically-improvable quantity to access amplitudes from real-time correlation functions. Furthermore, we use the spectral decomposition of correlation functions to provide empirical estimates of the order of magnitude of the size of volumes and times needed to study scattering amplitudes. 

First, we review the relevant concepts of scattering theory and the finite-volume formalism. The results presented in this section, which have been previously derived in the literature, only assume that the kinematics are such that only two-particle states can go on shell. 

\subsection{Infinite-volume formalism in 1+1D} \label{subsec:review:infvol}

First, we review the analytic properties of purely hadronic amplitudes ($\mathcal{M}$), transition amplitudes ($\mathcal{H}$), and Compton-like amplitudes ($\mathcal{T}$). The results presented here follow from Refs.~\cite{Briceno:2020vgp, Sherman:2022tco}. The only subtlety in relating the results here with what is in the literature is the fact that we will only be considering $1+1$D spacetime, while the literature normally focuses on amplitudes in $3+1$D spacetime. This results in the kinematic functions being different, but these were derived in our previous work~\cite{Briceno:2020rar}. 

 As detailed in Refs.~\cite{Briceno:2020vgp, Sherman:2022tco}, in a given kinematic region, all kinematic singularities in the amplitudes can be isolated exactly. As a result, one can write amplitudes in terms of known singular functions and generally unknown real functions that encode all of the dynamics. We first discuss the results for kinematics where only one intermediate channel can go on-shell, and we then quickly lift this assumption to allow any number of channels. For simplicity, we will always assume that the channels are composed of spinless identical bosons.  

\subsubsection{Single-channel case}
\label{sec:singleIV_sec}

Consider the center-of-momentum (c.m.) energy region $2m<\sqrt{s}<\sqrt{s}_{th}$, where $s$ is the Mandelstam variable, and $\sqrt{s}_{th}$ is the first unaccounted threshold. This could be a threshold associated with two or more particles, depending on the details of the theory and the channel. In this kinematic region, only the desired two-particle state can go on shell. Following the steps presented in Ref.~\cite{Briceno:2020vgp}, one can write the on-shell representation of the two-particle scattering amplitude in terms of the two-particle phase-space ($\rho$), and an unknown real function, referred to as the K matrix ($\mathcal{K}$),
  \begin{align}\label{eq:M}
\mathcal{M}(s) 
    = \frac{1}{\mathcal{K}(s)^{-1}-i\rho(s)}. 
  \end{align}
The phase space for two identical particles in $1+1$D is given by
  \begin{align}
\rho(s) = \frac{1}{8\sqrt{s}k^\star},
\label{eq:rho}
  \end{align}
where $k^\star=\sqrt{s/4-m^2}$ is the magnitude of the c.m. frame relative linear momentum between two identical particles. The square root in the definition of $k^\star$ is the one kinematic singularity present in the amplitude in this c.m. energy region. All other singularities, e.g. bound states and resonance poles, are dynamically generated by the K matrix.

In this same kinematic region, transition processes between one- and two-particle states involving a single current insertion can be described in terms of Eq.~\eqref{eq:M} and real-valued energy-dependent form factor function. Following the Lorentz decomposition of amplitudes, depending on the current considered, there might be multiple form factors. In this work, we will assume the current is a Lorentz scalar, which assures that there is only one energy-dependent form factor. 

To write a matrix element representation for these amplitudes, we need to define the current and the states. Let $\mathcal{J}(x)$ be a local scalar current defined at an arbitrary spacetime point $x$. The infinite-volume states for a single particle carrying  momentum $p=(\omega_\mathbf{p},\mathbf{p})$
are labeled $|p\rangle$. They have the standard relativistic normalization $\langle p| k \rangle = 2\omega_{\mathbf{p}} \, (2\pi) \delta(\mathbf{p}-\mathbf{k})$, where 
$\omega_\mathbf{p}=\sqrt{m^2+\mathbf{p}^2}$. The two-particle states, which can be constructed from the one-particle states, will be simply written as $|P;2\rangle$, where $P=(E,\mathbf{P})$ is the total spacetime-momentum of the system, and it is related to $s$ by $s=P^2$.

With this, we can write the transition amplitude for this case in terms of $\mathcal{M}$, and an energy-dependent form factor ($\mathcal{A}$), as~\cite{Briceno:2020vgp}
  \begin{align}
  \label{eq:HAM}
\mathcal{H}(s,Q^2)=\langle P;2\vert\mathcal{J}(0)\vert p\rangle = \mathcal{A}(s,Q^2)\mathcal{M}(s),
  \end{align}
where $Q^2=-q^2=-(P-p)^2$ is the so-called virtuality of the process and $q=P-p$ is the momentum transfer. The key point of Eq.~\eqref{eq:HAM} is that the physical singularities and consequently the phases of the transition amplitude are all encoded in $\mathcal{M}$. Note that this is equivalent to Watson's theorem~\cite{Watson:1952ji}.

Similarly, in this same kinematic region, the on-shell representation for Compton-like scattering amplitudes can be written in terms of $\mathcal{M}$ and purely real and smooth functions~\cite{Sherman:2022tco}. We only consider processes involving the previously introduced scalar currents and particles.  With this, we can write the on-shell projection for the Compton-like amplitude as~\cite{Sherman:2022tco}
  \begin{align}\label{eq:TwAMA}
\mathcal{T}(s,u,Q^2,Q^2_{if}) = w(s,u,Q^2,Q^2_{if}) 
    + \mathcal{A}(s,Q^2)\mathcal{M}(s)\mathcal{A}(s,Q^2_{if}) + [s\leftrightarrow u], 
  \end{align}
where $w$ is a new unknown smooth function, $p_f(p_i)$ is the momentum of the final(initial) single-particle state, and $q=(q^0,\mathbf{q})$ is the momentum of one of the currents, so $s=(p_f+q)^2$ and $u=(p_i-q)^2$ are the Mandelstam variables, with $Q^2=-q^2$ and $Q^2_{if}=-(p_f+q-p_i)^2$ being the virtualities of the two currents. The third term of Eq.~\eqref{eq:TwAMA}, $[s\leftrightarrow u]$ is obtained by reevaluating the second term in the r.h.s but with the Mandelstam variable $u$ and exchanging $Q^2\leftrightarrow Q^2_{if}$. The relation between $s$ and $u$ is detailed in Sec.~\ref{app:kinematics}. 

In what follows it will be necessary to consider a complementary representation of the Compton-like amplitude in terms of time-dependent matrix elements of the currents, 
  \begin{align}\label{eq:TME}
\mathcal{T}(s,u,Q^2,Q^2_{if}) 
    \equiv i \int d^2x~e^{iq \cdot x}
    \langle p_f \vert \mathbb{T}\{\mathcal{J}(x)\mathcal{J}(0)\} 
    \vert p_i \rangle,
  \end{align}
where the integral runs over all of spacetime and $\mathbb{T}$ indicates the time ordering of the currents. It is worth emphasizing that it is this presentation that is most amenable to real-time computations. In principle, one can envision constructing single-particle states. After doing so, one can evaluate such matrix elements by repeatedly inserting currents at different points in spacetime. Formally, the integral runs over an infinitely large spacetime. The fact that this is impossible in practice is the main focus of this work.

In what follows, we will make use of the relationship between $\mathcal{M}$, $\mathcal{H}$, and $\mathcal{T}$, which follows from the LSZ reduction formula. In particular, consider a current $\mathcal{J}$ that has the same quantum numbers as the single-particle state. Amplitudes involving these current insertions must have poles as functions of their virtuality at the mass of the particle, i.e. $Q^2\to-m^2$.

From LSZ, one can relate the residues of these poles with other physical amplitudes. For example, the residue of the $\mathcal{H}$ amplitude in the vicinity of $Q^2\sim-m^2$, is proportional to $\mathcal{M}$. In particular, 
  \begin{align}\label{eq:HLSZ}
\mathcal{M}(s) 
    = \lim_{Q^2\to-m^2} 
        \frac{(Q^2+m^2)}{\langle0\vert\mathcal{J}(0)\vert p\rangle}
        \mathcal{H}(s,Q^2)
  \end{align}
where $\langle0\vert\mathcal{J}(0)\vert p\rangle$ is the vacuum-to-one particle matrix element, or equivalently the decay constant of the single-particle state.  Equation~\eqref{eq:HLSZ}, in conjunction with Eq.~\eqref{eq:HAM}, implies that the $\mathcal{A}$ function satisfies 
  \begin{align}\label{eq:limA}
\lim_{Q^2\to-m^2}  
    \frac{(Q^2+m^2)}{\langle0\vert\mathcal{J}(0)\vert p\rangle} \mathcal{A}(s,Q^2)
    = 1.
  \end{align}

Similarly, one can recover the $\mathcal{H}$ function from the $\mathcal{T}$ amplitude, 
  \begin{align}\label{eq:HLSZT}
\mathcal{H}(s,Q^2) 
    = \lim_{Q^2_{if}\to-m^2}
        \frac{(Q^2_{if}+m^2)}{
            \langle p_f+q-p_i\vert\mathcal{J}(0)\vert 0\rangle}
        \mathcal{T}(s,Q^2,Q^2_{if}),
  \end{align}
and consequently also the $\mathcal{M}$ amplitudes from $\mathcal{T}$,
  \begin{align}\label{eq:MLSZT}
\mathcal{M}(s) 
    = \lim_{Q^2,Q^2_{if}\to-m^2}
        \frac{(Q^2+m^2)(Q^2_{if}+m^2)}{
            \langle0\vert\mathcal{J}(0)\vert q\rangle
            \langle p_f+q-p_i\vert\mathcal{J}(0)\vert 0\rangle}
        \mathcal{T}(s,Q^2,Q^2_{if}).
  \end{align}
These relations are used in Sec.~\ref{sec:LSZ} to construct an estimator for the two-particle scattering amplitude as well as the transition amplitude.

\subsubsection{Multiple-channel case}
\label{sec:MultiIV_sec}

The generalization of these identities to kinematics where any number of two-particle states can go on-shell is straightforward. For this scenario, we label the masses of the particles in the $a^{th}$ channel as $m_a$, with the lowest possible mass being $m_1$. With this, we can define the kinematic restriction for the following expressions as $2m_1 \leq \sqrt{s} < \sqrt{s_{th}}$, where once again $\sqrt{s_{th}}$ is the first unaccounted threshold. In Sec.~\ref{sec:InfT}, we will consider models where there are only two-particle states. As a result, it will effectively be the case that $\sqrt{s_{th}}\to\infty$. 

The equations above can be easily generalized by making the different building blocks either matrices or vectors in channel space\cite{Briceno:2012yi,Briceno:2014oea,Hansen:2012tf,Briceno:2014uqa}. For example, the phase space in Eq.~\eqref{eq:rho} is replaced by the diagonal matrix 
  \begin{align}
\rho_{ab}(s) =
\frac{\delta_{ab}}{8\sqrt{s}k^\star_a},
\label{eq:rhoab}
  \end{align}
where $k^\star_a=\sqrt{s/4-m_a^2}$. Meanwhile, $\mathcal{M}$ and $\mathcal{H}$ become a matrix and vector, respectively,  
  \begin{align}
  \label{eq:Mab}
\mathcal{M}_{ab}(s) &= 
    \left[ \mathcal{K}(s)^{-1} - i\rho(s) \right]_{ab}^{-1},\\
\label{eq:HAMab}
    \mathcal{H}_{b}(s,Q^2) &= 
    \mathcal{A}_{a}(s,Q^2)\mathcal{M}_{ab}(s),
  \end{align}
where in the last equality repeated indices are summed over. From these equations, it is clear that $\mathcal{K}$ and $\mathcal{A}$ are also a matrix and vector, respectively, in channel space. The elements of $\mathcal{M}$ in Eq.~\eqref{eq:Mab} describe purely hadronic scattering between two particles of types ``$a$'' and ``$b$''. Similarly, the elements of $\mathcal{H}$ describe the transition between a single-particle state coupled with the scalar current into two particles of type ``$b$''. For simplicity, we fix the single-particle state to be of type ``1''. As we discuss below, this assumption can be easily lifted to consider arbitrary scattering amplitudes. 

Irrespective of the kinematics, the Compton-like scattering amplitudes are by definition scalars in channel space. This is because their external states are fixed as one scans in energies. That said, the building blocks in Eq.~\eqref{eq:TwAMA} change as one allows for the kinematics of additional channels to open. Given Eqs.~\eqref{eq:Mab} and \eqref{eq:HAMab}, it is hopefully evident that the generalization of Eq.~\eqref{eq:TwAMA} for these kinematics is
  \begin{align}
  \label{eq:ComptonAnalytic}
\mathcal{T}(s,u,Q^2,Q^2_{if}) = 
    w(s,u,Q^2,Q^2_{if}) + 
    \mathcal{A}_a(s,Q^2) \mathcal{M}_{ab}(s) \mathcal{A}_b(s,Q^2_{if})+[s\leftrightarrow u]. 
  \end{align}
 Although this expression holds for any external one-particle states, here we will assume that the external particles are of type ``1''. In other words, the matrix-element representation shown in Eq.~\eqref{eq:TME} holds for arbitrary kinematics. The only restriction is that the momenta of the external particles satisfy $p_i^2 =p_f^2=m_1^2$. 

As in the single-channel case, one can use the LSZ reduction formula to go from $\mathcal{T}$ to $\mathcal{H}$, and from $\mathcal{H}$ to $\mathcal{M}$ as shown above. In what follows we will fix the current to only have the quantum number of one of the particles in channel 1. In other words, we give a label ``a'' to the single-particle states associated with the type of particle, $|k,a\rangle$, then 
\begin{align}
\langle k,a\vert\mathcal{J}(0)\vert 0\rangle
= \delta_{a1}\langle k,a\vert\mathcal{J}(0)\vert 0\rangle.
\end{align}

With this, we can write the relevant relations to obtain the $\mathcal{H}_1$ and $\mathcal{M}_{11}$ components of the corresponding amplitudes,
\begin{align}
\label{eq:HLSZ_11}
\mathcal{M}_{11}(s) 
    &= \lim_{Q^2\to-m^2_1} 
        \frac{(Q^2+m^2_1)}{\langle0\vert\mathcal{J}(0)\vert p,1\rangle}
        \mathcal{H}_1(s,Q^2),
\\
\label{eq:HLSZT_1}
\mathcal{H}_1(s,Q^2) 
    &= \lim_{Q^2_{if}\to-m^2_1}
        \frac{(Q^2_{if}+m^2_1)}{
            \langle p_f+q-p_i,1\vert\mathcal{J}(0)\vert 0\rangle}
        \mathcal{T}(s,Q^2,Q^2_{if}),
  \end{align}
which are simple generalizations of Eqs.~\eqref{eq:HLSZ} and \eqref{eq:HLSZT}, respectively. This puts tight constraints on the residue of $\mathcal{A}_b$ at the pole. Although $\mathcal{A}_b$ is generally a vector, at its pole, it must satisfy 
  \begin{align}
  \label{eq:limA_mc}
\lim_{Q^2\to-m^2_1}  
    \frac{(Q^2+m^2_1)}{\langle0\vert\mathcal{J}(0)\vert p,1\rangle} \mathcal{A}_b(s,Q^2)
    = \delta_{b1}.
  \end{align}

One can easily generalize this to obtain any components of the various amplitudes by choosing the external single-particle states and currents appearing in Eq.~\eqref{eq:TME} to have the quantum numbers of the desired external two-particle states. 

More generally, one can use this procedure to construct scattering amplitudes involving any number of external legs. For example, if one wants a purely hadronic amplitude involving $n/n'$ particles in the initial/final state, one can do this by evaluating matrix elements of $n+n'-2$ currents between an initial and a final single-particle state, where the currents are defined to have the quantum number of the desired particle.  One would then need to perform $n+n'-3$ Fourier transforms. From the resultant amplitude, one can isolate the residues from the $n+n'-2$ poles, which would be proportional to the $n\to n'$ hadronic amplitude.

\subsection{Finite-volume formalism in 1+1D} \label{subsec:review:finvol}

Having reviewed the infinite-volume on-shell representation of the scattering amplitudes, for both single- and multiple-channel systems, we now proceed to discuss the analogous finite-volume amplitudes $\mathcal{M}_L$, $\mathcal{H}_L$, and $\mathcal{T}_L$. Strictly speaking, scattering amplitudes are not well defined in a finite volume, but these are functions that, in a carefully defined infinite-volume limit, coincide with physical scattering amplitudes. The literature focuses on finite-volume amplitudes in 3+1D, see for example Refs.~\cite{Kim:2005gf, Briceno:2019opb, Briceno:2015csa, Baroni:2018iau, Briceno:2015tza}, but as we showed in Ref.~\cite{Briceno:2020rar}, it is straightforward to rewrite these results in 1+1D. The difference amounts to deriving some geometric functions for the dimensions of the considered spacetime. 

Before going over the existing formalism, it is important to summarize the key principles behind them. The main idea is that one can always isolate the power-law finite-volume effects exactly. These typically arise from intermediate multiparticle states going on-shell. Generally, power-law effects of finite-volume amplitudes can be determined non-perturbatively using an effective field theory representation skeleton expansion~\cite{Kim:2005gf}. The resulting expressions are correct up to the $\mathcal{O}(e^{-mL})$ errors, where $m$ is the mass of the lightest particle in the theory and $L$ is the spatial extent of the volume. This means that for sufficiently large volumes, these errors can be safely ignored, and we will do so in this work. Ultimately, the finite-volume amplitudes can be expressed in terms of finite-volume geometric functions and the infinite-volume scattering observables associated with the physical subprocesses, all of which were introduced in the previous section. Within the class of amplitudes and in the kinematic regions considered, only one type of these finite-volume geometric functions is required.

\subsubsection{Single-channel case}
\label{sec:singleFV_sec}

Let us begin by defining the finite-volume amplitudes for kinematics where only a single channel is open. We will use the same convention for labeling masses of the particle in a given channel that we introduced in the previous section. In other words, in the case where there is a single channel like here, we fix the particles to be identical with mass $m$. For the case with an arbitrary number of channels, we label the mass with the channel number. 
Similarly, we will keep the same assumption previously imposed on the c.m. energy. As a result, the only new feature in this section is the presence of a finite volume of size $L$. Imposing periodic boundary conditions, the linear momenta take discrete values $\mathbf{p}=\frac{2\pi\mathbf{d}}{L}$, with $\mathbf{d}$ being an integer. Here we will ignore effects associated with the temporal extent ($T$). Finite-$T$ effects will be covered in Sec.~\ref{sec:FiniteT}.

In this work, all power-law finite-volume effects can be encoded in a single kinematic function $(F)$, which depends on the total energy-momentum $P=(E,\mathbf{P})$ and $L$. This function is known exactly and can be derived by evaluating the difference between the finite- and infinite-volume $s$-channel two-particle loop. For two spinless identical bosons in 1+1D, this function is given by the sum-integral difference,
   \begin{align}
F(E,\mathbf{P},L) 
&= 
    \lim_{\epsilon\to0^+} \frac{1}{2}
    \left[ \frac{1}{L}\sum_{\mathbf{p}} 
        - \int\frac{d\mathbf{p}}{2\pi} \right]
    \frac{1}{2\omega_{\mathbf{p}}}
    \frac{1}{(P-p)^2-m^2+i\epsilon},
\nonumber \\
&= 
    i\rho(s)+\rho(s)\cot
    \left(\frac{1}{2}\left(\gamma Lk^\star-\mathbf{d}\pi\right)
    \right), \label{eq:F}
  \end{align} 
where $\gamma=E/\sqrt{s}$, and $\mathbf{d}=\frac{L\mathbf{P}}{2\pi}$. This form of $F$ is equivalent to Eq.~(22) in Ref.~\cite{Briceno:2020rar}.\footnote{Using the expressions and notation of Eq. (22) in Ref.~\cite{Briceno:2020rar}, the equivalence with Eq.~\eqref{eq:F} can be seen by first noting that for identical particles $L\gamma \omega_q^\star \beta  =L  \mathbf{P} \omega_q^\star/E^\star = \pi \mathbf{d}$. One can then use standard trigonometric identities to show the equality.} 

Given $F$ and $\mathcal{M}$, one can find a closed-form expression for the finite-volume analog of $\mathcal{M}$, thus 
  \begin{align}\label{eq:ML}
\mathcal{M}_L(E,\mathbf{P}) 
    = \frac{1}{\mathcal{M}(s)^{-1}+F(E,\mathbf{P},L)}.
  \end{align}
One key observation is that the finite-volume function depends on the energy and momentum of the system, while the infinite-volume amplitude is a Lorentz scalar depending only on $s$. This dependence will be key in what follows. We choose to emphasize the $L$ dependence in the name of the function, rather than within its arguments, to clearly distinguish it from the infinite-volume amplitude. 

Similarly, one can define the finite-volume analog of the transition amplitude. Considering the same scenario previously discussed, where the current and all the particles involved are scalars, the finite-volume transition amplitude can be compactly written as
  \begin{align}\label{eq:HL}
\mathcal{H}_L(E,\mathbf{P},Q^2)
    &= \mathcal{A}(s,Q^2)\mathcal{M}_L(E,\mathbf{P}). 
  \end{align}
It is interesting to note that, as shown in Eq.~\eqref{eq:HAM} in the infinite-volume size, the singularities of the transition amplitude $\mathcal{H}$ are given by those of $\mathcal{M}$. In a finite volume, we see that the power-law finite-volume effects of $\mathcal{H}_L$ are given by $\mathcal{M}_L$.

Finally, as derived in Ref.~\cite{Briceno:2019opb}, the corresponding finite-volume amplitude for Compton-like scattering is 
  \begin{align}\label{eq:TL}
\mathcal{T}_L(p_f,q,p_i) 
    &= w(s,u,Q^2,Q^2_{if})+\mathcal{A}(s,Q^2)\mathcal{M}_L(E_f + q^0,\mathbf{p}_f+\mathbf{q})\mathcal{A}(s,Q^2_{if})
    +[s\leftrightarrow u],
  \end{align} 
where the momenta $p_f$, $q$, and $p_i$ and $[s\leftrightarrow u]$ are the same as in Eq.~\eqref{eq:TwAMA}. Once again, there is a close parallel between the infinite-volume amplitude $\mathcal{T}$, in Eq.~\eqref{eq:TwAMA}, and the finite-volume amplitude counterpart. Basically, one recovers the latter from the former by making the simple replacements $\mathcal{M}\to\mathcal{M}_L$.

We started this section by stating that in a finite volume, scattering amplitudes are not well defined. One is na\"ively tempted to use Eqs.~\eqref{eq:ML}~to~\eqref{eq:TL} as a working definition of finite-volume amplitudes. As discussed in detail in Ref.~\cite{Briceno:2020rar}, the finite-volume amplitudes are purely real sums of poles. Meanwhile, infinite-volume amplitudes are complex-valued functions with branch-cut singularities. This disparity is a manifestation of the previous statement. 

Just as in the infinite-volume case, one can write a matrix-element representation of $\mathcal{T}_L$, 
  \begin{align}\label{eq:TLME_2}
\mathcal{T}_L(p_f,q,p_i)
    \equiv i2\sqrt{\omega_{\mathbf{p}_f}\omega_{\mathbf{p}_i}}L
    \int_{-T/2}^{T/2} dt \int_0^L d\mathbf{x}~e^{iq \cdot x-|t|\epsilon}
    \langle p_f,L \vert \mathbb{T}\{\mathcal{J}(x)\mathcal{J}(0)\} 
    \vert p_i,L \rangle\bigg|_{\epsilon =0},
  \end{align}
where $\vert p,L\rangle$ is a single-particle state in a finite volume with  momentum $p$, normalized as $\langle p,L\vert q,L\rangle=\delta_{\mathbf{pq}}$. Note, we have introduced an $\epsilon$ to regulate the integral, which is taken to zero after integration. In Sec.~\ref{subsec:review:estimators}, we will make this $\epsilon$ explicit in the definition of the estimators. This is the same definition presented in Ref.~\cite{Briceno:2020rar}, except that we have truncated the time extent of the integral. This reflects the fact that in practice the number of measurements for which one can evaluate these matrix elements will have to be finite. As a result, the time will have to be truncated. 

The fact that the finite-volume amplitudes are real for real energies can be seen from the definitions of $\mathcal{M}$ and $F$ from Eqs.~\eqref{eq:M}~and~\eqref{eq:F}. The poles in $\mathcal{M}_L$ coincide with the spectrum of the two-particle states in a finite volume. With this in mind, for a given boost ($\mathbf{P}$) and $L$, the spectrum of energies of such states is given by 
  \begin{align}\label{eq:Luscher1D}
\mathcal{M}(s_n)^{-1}+F(E_n,\mathbf{P},L)=0,
  \end{align}
where $\sqrt{s_n}=\sqrt{E_n^2-\mathbf{P}^2}$. Equation~\eqref {eq:Luscher1D} is the 1+1D, single channel version of the so-called L\"uscher quantization condition~\cite{Briceno:2020rar}.

  \subsubsection{Multiple-channel case}
\label{sec:MultiFV_sec}

Having seen in Sec.~\ref{sec:MultiIV_sec} how the analytic expression for single-channel scattering amplitudes can be easily generalized to kinematics where an arbitrary number of channels can go on shell, it is not hard to see how this can be done for finite-volume amplitudes. In short, this can be done by upgrading every building block to either a vector or matrix in the number of open channels.

The only new building block is the finite-volume $F$ function. When multiple channels are open, the $F$ function becomes a diagonal matrix over these channels, with matrix elements
  \begin{align}
F_{ab}(E,\mathbf{P},L) \label{eq:Fab}
    &= i\rho_{ab}(s)
    +\rho_{ab}(s)\cot
    \left(\frac{1}{2}\left(\gamma Lk^\star_a-\mathbf{d}\pi\right)
    \right). 
  \end{align}
In this case the exponentially suppressed finite-volume corrections scale, at worst, as $\mathcal{O}( e^{-m_1L} )$, with $m_1$ being the mass of the lightest particle. 

With this, we can immediately write the generalization of the various finite-volume amplitudes, which we summarize here
  \begin{align}
\mathcal{M}_{Lab}(  E,\mathbf{P}) 
    &= [\mathcal{M}(s)^{-1}+F(E,\mathbf{P},L)]^{-1}_{ab},\\
\label{eq:TLab}
\mathcal{H}_{Lb}( E,\mathbf{P},Q^2)
    &=\mathcal{A}_a(s,Q^2)\mathcal{M}_{Lab}(E,\mathbf{P}), \\
\mathcal{T}_{L}(p_f,q,p_i)
    &= w(s,u,Q^2,Q^2_{if}) 
    + \mathcal{A}_a(s,Q^2)\mathcal{M}_{Lab}(E_f +\omega,\mathbf{P})\mathcal{A}_b(s,Q^2_{if})
    + [s\leftrightarrow u].
  \end{align} 

The previous statements made about the singularity of the finite-volume amplitudes persist even when an arbitrary number of channels can go on shell. In particular, the singularities of these are given by the more general form of the L\"uscher quantization condition~\cite{Hansen:2012tf, Briceno:2012yi}, 
  \begin{align}
\det\left[ 1 + \mathcal{M}(s_n)F(E_n,\mathbf{P},L) \right]=0,
  \end{align}
where the determinant is taken over the channel space.

  \subsection{Review of estimators} 
  \label{subsec:review:estimators}

In Ref.~\cite{Briceno:2020rar} we presented a proposal for recovering the infinite-volume Compton-like amplitude from real-time finite-volume correlation functions. This procedure consists of constructing an estimator for the amplitude  $(\overline{\mathcal{T}})$ based on three key ingredients: 
\begin{itemize}
    \item \textit{Finite $\epsilon$-prescription},
    \item \textit{Lorentz invariance},
    \item \textit{Energy binning/averaging},
\end{itemize}
each of which we describe below. From this, we will use the LSZ formalism to define estimators for the transition ($\overline{\mathcal{H}}$) and hadronic ($\overline{\mathcal{M}}$) amplitudes.

 Parting from the finite-volume matrix elements in Eq.~\eqref{eq:TLME_2}, we can proceed to review the proposed estimator. First, we need to introduce a small but finite imaginary term into the integral that dampens the integrand as $|t|$ becomes large,  
  \begin{align}\label{eq:TLME_eps}
\mathcal{T}_L(p_f,q,p_i, \epsilon)
    \equiv i2\sqrt{\omega_{\mathbf{p}_f}\omega_{\mathbf{p}_i}}L
    \int_{-T/2}^{T/2} dt \int_0^L d\mathbf{x}~e^{iq \cdot x-|t|\epsilon}
    \langle p_f,L \vert \mathbb{T}\{\mathcal{J}(x)\mathcal{J}(0)\} 
    \vert p_i,L \rangle.
  \end{align}
This is equivalent to shifting the energy of the finite-volume amplitude slightly away from the real axis, where the finite-volume poles lie, thereby softening these singularities.~\footnote{Note, in Eqs.~\eqref{eq:TLME_2}~and~\eqref{eq:TLME_eps} we use the same symbol to describe the $\epsilon$-independent and $\epsilon$-dependent amplitude. }

 It is important to emphasize that by accessing time-dependent matrix elements and performing Fourier transforms, as dictated by Eq.~\eqref{eq:TLME_eps}, one has complete analytic control on the values of $q$ that would be accessed, including time-like values. This is a necessary condition to analytically continue to the single-particle poles of the amplitudes and apply the LSZ reduction formula.

Analytically continuing the finite-volume amplitudes into the complex plane allows for a more direct comparison between $\mathcal{T}_L$ and $\mathcal{T}$. In fact, as shown in Ref.~\cite{Briceno:2020rar}, the infinite-volume amplitude can in principle be recovered from the ordered limits 
  \begin{align} \label{eq:dlimit}
\mathcal{T}(s,u,Q^2,Q^2_{if}) 
    =\lim_{\epsilon\to0}\lim_{L\to\infty}
        \mathcal{T}_L(p_f,q,p_i,\epsilon).
  \end{align}
However, implementing this limit would require prohibitively large volumes for reasonable estimates of the infinite-volume amplitudes.

The next step is to exploit the symmetry of the infinite-volume amplitude, which is a Lorentz scalar that can only depend on kinematic invariants. In contrast, the finite-volume amplitude is not a Lorentz scalar. However, for a given volume there is a number of finite-volume momenta $(\mathbf{p}_f,\mathbf{q},\mathbf{p}_i)$ that can be mapped to the target invariants ($Q^2$, $Q^2_{if}$, $s$, $u$) of the desired amplitude. The key observation presented in Ref.~\cite{Briceno:2020rar}, is that by averaging over the finite-volume kinematics that are approximately close to the target variables, the finite-volume artifacts are significantly dampened.

Let us focus our attention on a set of external momenta that results in the c.m. energy being within the kinematic region of interest, $2m<\sqrt{s'}<\sqrt{s}_{th}$. Within this set, there is a smaller set that would have virtualities $Q'^2$ and $Q'^2_{if}$ close to their targeted values. With this in mind, we average the finite-volume amplitudes that have momenta within the set of kinematic points that satisfy the following constraints
  \begin{align}
\left\vert Q^2-Q'^2\right\vert&<\Delta_{Q^2} , \hspace{1cm}
\left\vert Q^2_{if}-Q'^2_{if} \right\vert < \Delta_{Q^2},\hspace{1cm} 
\left\vert \sqrt{s} -\sqrt{s'}\right\vert < \Delta_{\sqrt{s}},
\label{eq:binning}
  \end{align}
where $\Delta_{Q^2}$ and $\Delta_{\sqrt{s}}$ are generally small quantities whose exact value may in general depend on the dynamics of the system.

This averaging has two effects. First, it effectively constructs wave packets with energies centered around $\sqrt{s'}$. Second, it enhances the symmetry of the resultant estimator, which further dampens the finite-volume effects.  

Letting $\mathcal{N}$ be the total number of kinematic points that satisfy Eq.~\eqref{eq:binning}, the proposed estimator for the  Compton-like amplitude is defined as,
  \begin{align}\label{eq:binnedT}
\overline{\mathcal{T}} (s,u,Q^2,Q^2_{if}) 
    = \frac{1}{\mathcal{N}} 
      \sum_L\sum_{\mathbf{p}_f,\mathbf{q},\mathbf{p}_i}
      \mathcal{T}_L(p_f,q,p_i,\epsilon).
  \end{align}
  
As mentioned above, we can use this Compton-like estimator in conjunction with the LSZ reduction formula to define estimators for transition and purely hadronic amplitudes, $\overline{\mathcal{H}}$ and $\overline{\mathcal{M}}$, respectively. For the transition amplitude, we can use Eqs.~\eqref{eq:HLSZT} to define its estimator, 
  \begin{align}\label{eq:estimatorH}
\overline{\mathcal{H}} (s,Q^2) 
    = \frac{1}{\mathcal{N}} 
      \sum_L\sum_{\mathbf{p}_f,\mathbf{q},\mathbf{p}_i}
      \frac{(Q'^2_{if}+m^2)}{\langle p_f+q-p_i\vert\mathcal{J}(0)\vert0\rangle}
      \mathcal{T}_L(p_f,q,p_i,\epsilon),
  \end{align}
where the average is over the set of kinematic points close to the on-shell value $Q^2_{if}\sim-m^2$ within the criteria defined above.

Similarly, by also fixing the target virtuality of the outgoing current as $Q^2\sim-m^2$, an estimator for the purely hadronic scattering amplitude can be constructed as 
  \begin{align}\label{eq:estimatorM}
\overline{\mathcal{M}} (s) 
    = \frac{1}{\mathcal{N}} 
\sum_L\sum_{\mathbf{p}_f,\mathbf{q},\mathbf{p}_i}
      \frac{(Q'^2+m^2)(Q'^2_{if}+m^2)}{\langle 0\vert\mathcal{J}(0)\vert q\rangle\langle p_f+q-p_i\vert\mathcal{J}(0)\vert0\rangle}
      \mathcal{T}_L(p_f,q,p_i,\epsilon).
  \end{align}
Notice that in both cases the pole subtraction is made in terms of the parameters $Q'^2$ or $Q'^2_{if}$, determined from given kinematic variables, and not in terms of the target virtualities. In other words, $Q'^2$ and $Q'^2_{if}$ are evaluated inside the sum. Empirically we have found that this definition improves the estimator for the residue at the pole. This could be interpreted as each element in the sum having a well-defined limit as one approaches the pole.

\section{ 
Numerical investigation of Estimators for amplitudes assuming infinite time extents }\label{sec:InfT}

As reviewed in Sec.~\ref{subsec:review:finvol}, the finite-volume Compton-like scattering amplitude, which is used to construct the estimators, can be parametrized using a finite number of real-valued functions, $\mathcal{K}$, $\mathcal{A}$, and $w$. To provide estimates of the resources needed for real-time calculations, we resort to using parameterizations of these functions. This allows us to compare finite-volume estimators to the infinite-volume amplitudes.  

Section~\ref{subsec:parz} presents our parametrization of the kinematic functions. In Sec.~\ref{sec:LSZ} we test the estimators $\overline{ \mathcal{ H}}$ and $\overline{\mathcal{M}}$ for systems with a single coupling channel. Then in Sec.~\ref{sec:multicc}, we test the estimators for systems with multiple coupling channels. This last case is considered only for $\overline{\mathcal{ T}}$ and $\overline{\mathcal{M}}$ since the implementation for $\overline{\mathcal{H}}$ can be thought of as a middle step in LSZ reducing Compton-like amplitudes to purely hadronic amplitudes.

Overall, we have observed that a numerical implementation of the LSZ reduction formula in the presence of any number of coupling channels does not require us to use volumes of a different order than those considered in Ref.~\cite{Briceno:2020rar} while keeping a good correspondence with the physical amplitudes. 

Throughout this section, we only consider the $s$-channel contribution to the Compton-like scattering amplitude. As we discuss in Sec.~\ref{sec:FiniteT}, including the $u$-channel contribution for these amplitudes is straightforward, and it does not change the estimates for the resources needed.

\subsection{Parametrization considered}\label{subsec:parz}

We have used different models to test the effectiveness of the estimators in recovering the infinite-volume amplitudes. To set these models we use a reasonably flexible parametrization of the K matrix which is smooth up to a simple pole singularity, which results in a resonant amplitude. Generally, the K matrix is a symmetric matrix over channels, and we consider the following parametrization for each element,
  \begin{align} \label{eq:kmat}
\mathcal{K}_{ab}(s) = 
    m_a m_b k_1^{\star2} 
    \left( 
        \frac{g_a g_b}{m_R^2-s} + h^{(0)}_{ab}(s)
    \right),
  \end{align}
where $g_a$ and $m_R$ are constants, and $h^{(0)}_{ab}$ is a polynomial in $s$. The overall $k^{\star 2}_1$ factor in Eq.~\eqref{eq:kmat} ensures that the K matrix vanishes at the first threshold. 

For $\mathcal{A}_a$ we consider classes of parameterizations of the form
   \begin{align}\label{eq:Aa}
\mathcal{A}_a(s,Q^2) = 
    \frac{1}{h_a^{(2)}(Q^2)}
    \left(
        \frac{h_a^{(1)}(Q^2)}{Q^2+m^2} + h_a^{(3)}(s,Q^2)
    \right),
  \end{align}
where the $h_a^{(i)}$ are polynomials in $s$ and/or $Q^2$. For simplicity, we consider $Q^2_{if}=Q^2$. In general, one can choose $Q^2$ and $Q^2_{if}$ to take on any independent values. 

Given we are just modeling amplitudes, there is flexibility in the parameterizations that can be considered for the $h_a^{(i)}$ polynomials. An important condition in what follows is the fact that $\mathcal{A}_a$ must satisfy the limit placed by the LSZ reduction formula, see Eq.~\eqref{eq:limA_mc}. In other words, the residue at the pole must be proportional to a Kronecker $\delta$ function. Since we will only consider the case where the external particles are of type ``1'', this implies that $h_a^{(1)}\propto \delta_{a1}$.     

As previously emphasized in Ref.~\cite{Briceno:2020rar}, the source of tension between finite- and infinite-volume amplitudes is due to the difference between their singular structure. Given this and the fact that the function $w$ is generally a smooth function in the kinematic region considered, we will fix it to zero. This condition, along with the parametrizations in Eqs.~\eqref{eq:kmat}~and~\eqref{eq:Aa} set the models used to test the estimators. 

For systems with multiple coupling channels, each of these is defined through channel-space tuples of coupling constants $\mathbf{g}$ and masses $\mathbf{m}$. If only single values of $g$ and $m$ are given, then all of the kinematic equations are assumed to be scalar objects and a single coupling channel is being considered.

\subsection{ 
Estimators for single-channel systems}
\label{sec:LSZ}

\begin{figure}[t]
\centering
\includegraphics[width=0.975\linewidth]{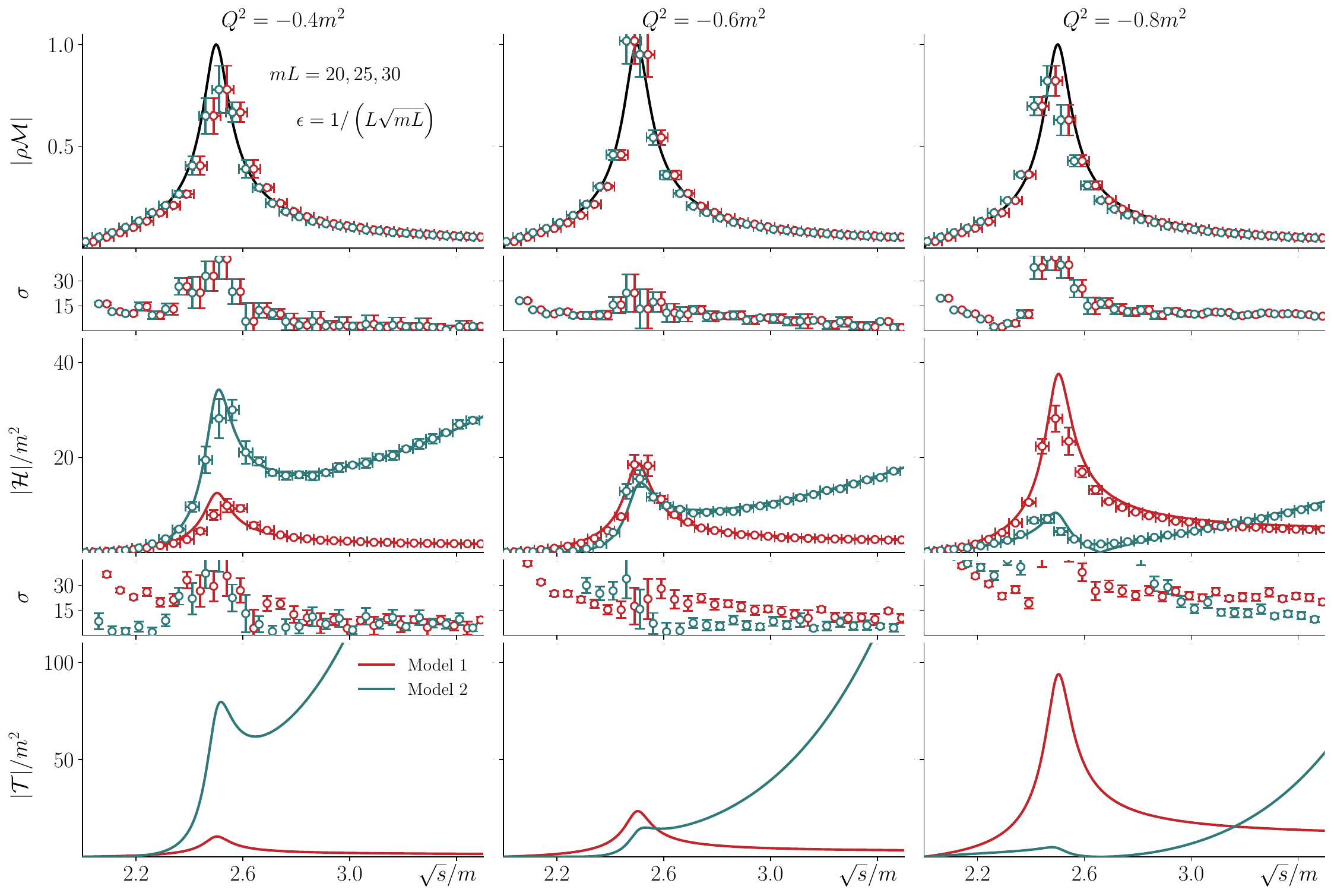}
\caption{ From top to bottom: purely hadronic, transition, and Compton-like amplitudes. The definition of $\sigma$ is given in Eq.~\eqref{eq:sigmaL}. The dots represent the numerical results for the estimators while the solid lines are the infinite-volume amplitudes. \textit{Model 1} is displayed in red with a slight offset to the right. \textit{Model 2} is displayed in blue with a slight offset to the left. Both cases are detailed in the text. From left to right, the estimators are computed for virtualities approaching the limit $Q^2\to-m^2$, with $Q^2_{if}=Q^2$. }  
\label{Fig:ModelsLSZ}
\end{figure}

In our investigation, we considered a broad class of examples. Here we show results for $\overline{\mathcal{H}}$ and $\overline{\mathcal{M}}$ for two illustrative example cases defined as:
  \begin{itemize}
      \item \textit{Model 1)} $h^{(1)}(Q^2)=0.5m^2$, $h^{(2)}(Q^2)=1$, and $h^{(3)}(s,Q^2)=0$,
      \item \textit{Model 2)} $h^{(1)}(Q^2)=m^2-Q^2$, $h^{(2)}(Q^2)=5Q^2/m^2$, and $h^{(3)}(s,Q^2)=1-0.2s^2/m^4$. 
  \end{itemize}
In both cases, the K matrix is set by $g=2.5$, $m_R=2.5m$, and $h^{(0)}=0$. For both cases, the pole of the $\mathcal{A}$ function corresponds to a particle of mass $m$, which assures that both the transition and Compton-like amplitudes have single-particle poles. This also implies that for these models, the hadronic amplitudes are the same while the transition and Compton-like amplitudes are different.  

The volumes considered in the boost averaging are $mL=20,25,30$. We consider any kinematics capped within $\Delta_{Q^2}=0.05m^2$ and $\Delta_{\sqrt{s}}=0.05m$ as in Eq.~\eqref{eq:binning}. To replicate the effect of the infinite-volume limit, Eq.~\eqref{eq:dlimit}, the $\epsilon$-prescription is tied to the volume size by $\epsilon=1/(L\sqrt{mL})$. These choices are somewhat arbitrary since there are many other choices in binning and $\epsilon$ that would yield comparable results. 

The numerical results of the estimators for both cases are presented in Fig.~\ref{Fig:ModelsLSZ}. The top panels show the estimator $\overline{\mathcal{M}}$ along with the corresponding infinite-volume amplitude $\mathcal{M}$. The middle panels show the equivalent result for $\overline{\mathcal{H}}$. For each quantity, we show the mean value as well as the error on the mean. The horizontal errors represent the width of the c.m. energy bins used to define the estimator. 

Below each panel, we show a measure of the systematic error of amplitude. For the hadronic amplitude it is defined as
\begin{align}\label{eq:sigmaL}
\sigma(s)\equiv
   100\times \frac{\left\vert 
        \overline{\mathcal{M}}(s)
        - 
        \mathcal{M}(s)
    \right\vert}{\mathcal{M}(s)}
    .
  \end{align}
Similar definitions are used for $\mathcal{H}$
and $\mathcal{T}$.  This measures the systematic deviation of the estimator away from the value of the target amplitude. In the figures, we also show the propagated statistical uncertainty onto the definition of $\sigma$. 

This provides a useful visualization to compare the systematic versus statistical error of an estimator. For example, if the error on $\sigma$ is greater than or equal to its mean value, one can conclude that the physical amplitude is being recovered within one standard deviation of the quoted statistical error. Otherwise, the opposite is true.

The bottom panels show the infinite-volume Compton-like amplitude for each case. Different values of $Q^2$ approaching the on-shell limit from the left, $Q^2\to-m^2$, are considered across the columns. 

\begin{figure}[t]
\centering
\includegraphics[width=0.985\linewidth]{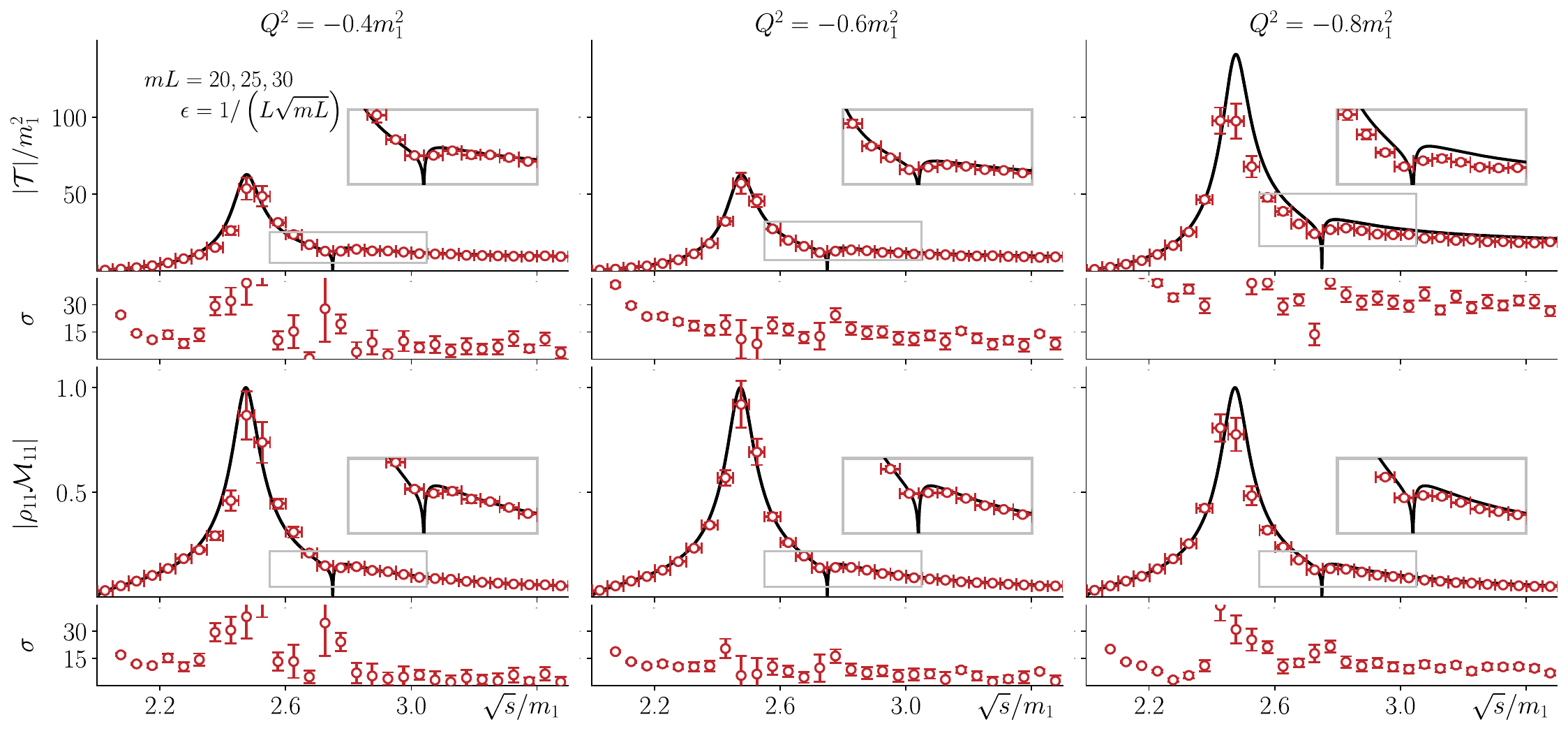} \\[3mm]
\caption{Compton-like (top) and hadronic $\mathcal{M}_{11}$ (bottom) amplitudes for \textit{Model 1}, as described in the text. There are two kinematically open channels corresponding to masses defined by $\mathbf{m}/m_1=(1,1.375)$. The different symbols are the same as in Fig.~\ref{Fig:ModelsLSZ}. The region where the second channel opens up is highlighted in the inset. In all cases we fix $Q^2_{if}=Q^2$. }  
\label{Fig:MCC2}
\end{figure}

Broadly speaking, the results can be summarized as follows. Away from the singularities of the amplitude, i.e. the threshold and the dynamical enhancement driven by a nearby resonance, the estimators reproduce the amplitudes for the different models relatively well. As the virtuality of the current gets increasingly close to the pole, the agreement slightly worsens. This is consistent with what was presented in Ref.~\cite{Briceno:2020rar}, where it was shown that at the pole the hadronic amplitudes will be systematically off. As a result, we find that the convergence improves if one averages over a set of points around the single-particle pole. Finally, as expected, following this procedure we recover the same hadronic amplitudes for the two models.

\subsection{ Estimators for coupled-channel systems}
\label{sec:multicc}

We have also considered a broader set of parameterizations involving coupled-channel systems. The two illustrative examples we report involve two and four open channels set by:
  \begin{itemize}
      \item \textit{Model 1)} $\mathbf{m}/m_1=(1,1.375)$,  $\mathbf{g}=(2.5, 1.25)$, and $\mathbf{h}^{(1)}(Q^2)/m^2_1=(1,0)$.
      \item \textit{Model 2)} $\mathbf{m}/m_1=(1,1.3,1.4,1.5)$, $\mathbf{g}=(2.5, 1.5, 1.25, 0.985)$, and $\mathbf{h}^{(1)}(Q^2)/m^2_1=(1,0,0,0)$. 
  \end{itemize}
In both cases we have $h^{(2)}_a=2Q^2/m^2$, $h^{(3)}_a=0$,  $h^{(0)}_{ab}=0$. For simplicity, all other parameters are fixed to be the same as above, and the incoming and outgoing target virtualities are kept equal, $Q^2_{if}=Q^2$.  

\begin{figure}[t]
\centering
\includegraphics[width=0.985\linewidth]{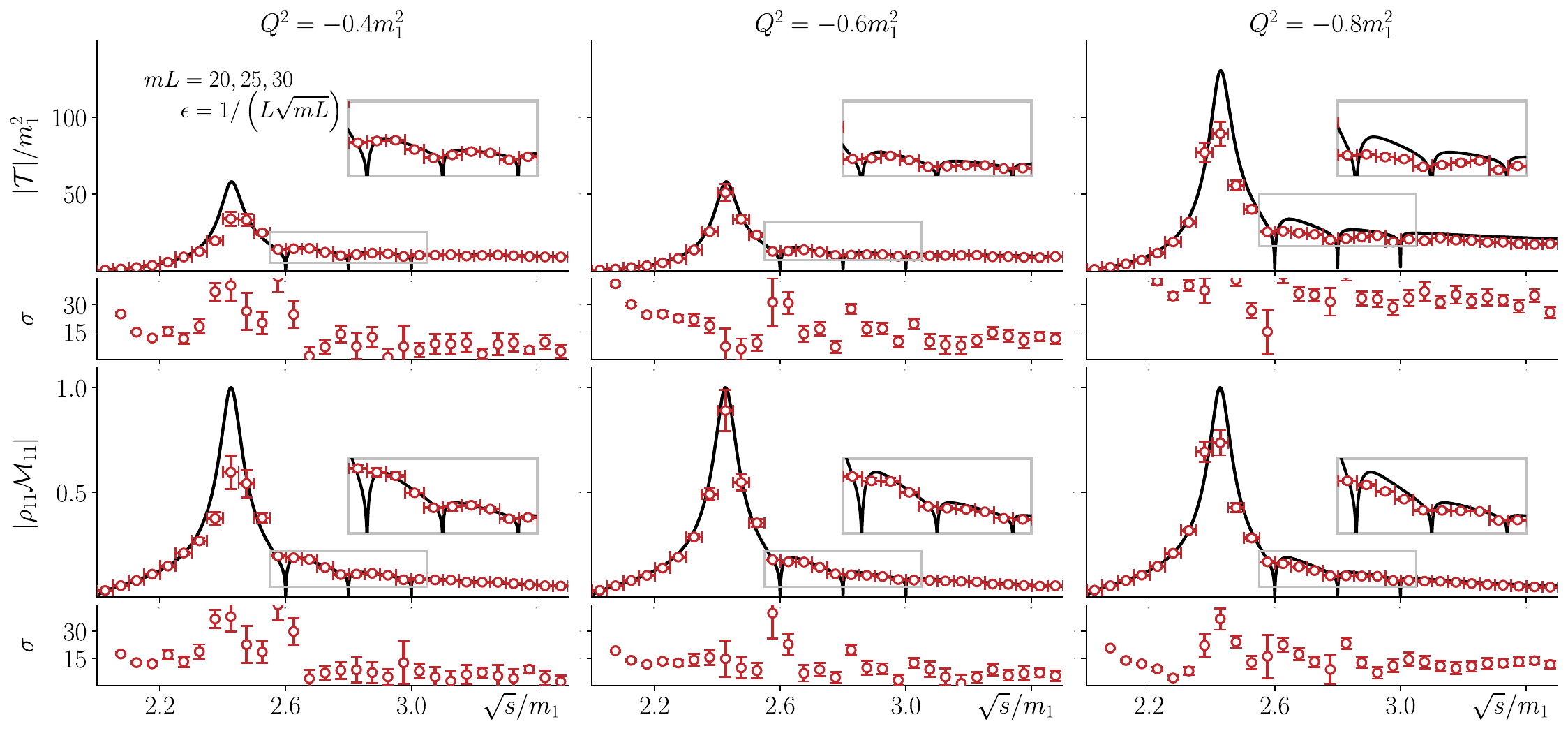} 
\caption{The same quantities as in Fig.~\ref{Fig:MCC2} for the second coupled-channel model, defined in the text as \textit{Model 2}. This has four open kinematic channels with masses defined by $\mathbf{m}/m_1=(1,1.3,1.4,1.5)$.
}  
\label{Fig:MCC4}
\end{figure}

The estimators for purely hadronic and Compton-like scattering amplitudes in the presence of multiple open coupling channels are presented in Fig.~\ref{Fig:MCC2} and Fig.~\ref{Fig:MCC4}. 
Note, $\mathcal{M}$ is now a matrix over channels, and we only show the results for $\mathcal{M}_{11}$. 

The conclusion from these figures is the same as for the single-channel case. Namely, for the setup considered the estimators do a reasonable job of reproducing the amplitudes. The main discrepancy lies near the various singularities, where the estimator struggles to faithfully reproduce the amplitudes. In these scenarios, we have more threshold singularities, and as can be seen from Fig.~\ref{Fig:MCC2} and Fig.~\ref{Fig:MCC4}, the estimators seem to struggle to resolve these kinematic singularities. 

As in the single-channel case, as one approaches the single-particle pole in the Compton-like amplitude, there seems to be a \emph{golden window} for being able to reasonably reconstruct the hadronic amplitude. If one gets too far or too close to the pole, one introduces larger systematic errors. Nevertheless, the hadronic amplitudes seem to be well-constrained for most kinematic points for the models considered.

As a result, it is safe to conclude that the proposed estimators presented in Ref.~\cite{Briceno:2020rar} work equally well independent of the number of open channels present. 

\section{The role of finite-time effects} \label{sec:FiniteT}

So far, we have discussed the estimators built under the assumption that $T\to\infty$. In this section, we lift this assumption and show the effects that finite-$T$ conditions have on the estimator $\overline{\mathcal{T}}$.  

\subsection{A spectral representation for Compton-like amplitudes}\label{subsec:Spectral}

To investigate the $T$ dependence of the estimators we begin by rewriting Eq.~\eqref{eq:TLME_eps} using the spectral decomposition of the matrix elements and evaluating the integral over time analytically. In the models we have considered the local current exclusively couples the external single-particle to two-particle states. With this in mind, we will need to insert a complete set of finite-volume two-particle states between the two currents. For example, for $t>0$, the matrix element takes the form
  \begin{align}\label{eq:decomposition}
\langle p_f,L \vert \mathcal{J}(x)\mathcal{J}(0) \vert p_i, L \rangle
    = \sum_{(E_n,\mathbf{P})} e^{ i(p_f-p_n)\cdot x }
    \langle p_f,L \vert \mathcal{J}(0) \vert P_n,L;2\rangle 
    \langle P_n,L;2\vert\mathcal{J}(0) \vert p_i, L \rangle, 
  \end{align}
where $\vert P_n,L;2\rangle$ is a finite-volume state with the quantum numbers of two-particles with total energy-momentum $P_n=(E_n,\mathbf{P})$. These states are normalized to unity, $\langle P_n,L;2  \vert P_{n'}',L;2\rangle =\delta_{n,n'}\delta_{\mathbf{P},\mathbf{P}'}$. The sum in Eq.~\eqref{eq:decomposition} goes over the energy levels $E_n$ satisfying the L\"uscher quantization condition for a given boost $\mathbf{P}$, see Eq.~\eqref{eq:Luscher1D}.  

Using Eq.~\eqref{eq:decomposition} for the time-dependent matrix elements and inputting it into the $t>0$ contribution of the integral appearing in Eq.~\eqref{eq:TLME_eps}, one can evaluate the two integrals to find,
  \begin{align}\label{eq:t_pos}
\nonumber
&i2\sqrt{\omega_{\mathbf{p}_f}\omega_{\mathbf{p}_i}}L
\int_0^{T/2}dt\int_0^L d\mathbf{x}~e^{iq \cdot x-t\epsilon}
\langle p_f,L \vert \mathcal{J}(x)\mathcal{J}(0) \vert p_i, L \rangle \\
    &\qquad = 2\sqrt{\omega_{\mathbf{p}_f}\omega_{\mathbf{p}_i}}L^2 \sum_{P_n} 
       \frac{ e^{i(q^0+\omega_{\mathbf{p}_f}-E_n)T/2-\epsilon t} - 1 }{ 
              q^0+\omega_{\mathbf{p}_f}-E_n+i\epsilon }
        \langle p_f,L\vert\mathcal{J}(0)\vert P_n,L;2 \rangle 
        \langle P_n,L;2 \vert \mathcal{J}(0) \vert p_i, L \rangle
        \delta_{\mathbf{p}_f+\mathbf{q},\mathbf{P}},  
  \end{align}
where $\delta_{\mathbf{p}_f+\mathbf{q},\mathbf{P}}$ is a Kronecker delta that imposes conservation of linear momenta and fixes $\mathbf{P}=\mathbf{p}_f+\mathbf{q}$. An equivalent expression can be written for $t<0$.

The matrix elements coupling one- and two-particle finite-volume states can be written in terms of the infinite-volume transition amplitudes and a multiplicative factor, generally known as the Lellouch-L\"uscher factor~\cite{Lellouch:2000pv,Briceno:2014uqa,Briceno:2015csa},
  \begin{align}\label{eq:2to1ME}
\left|\langle p_f,L\vert\mathcal{J}(0)\vert P_n,L;2\rangle \right|
    &= \sqrt{ \frac{ |\mathcal{R}(E_n,\mathbf{P},L)| }{2\omega_{\mathbf{p}_f}} }
    \frac{ |\mathcal{H}(s_n,-(p_f-P_n)^2)| }{L}
     \\
  \ &= \sqrt{ \frac{ |\mathcal{R}(E_n,\mathbf{P},L)| }{2\omega_{\mathbf{p}_f}} }
    \frac{ |\mathcal{A}(s_n,-(p_f-P_n)^2)\mathcal{M}(s_n)| }{L},
  \end{align}
where in the second equality we have used the definition of $\mathcal{H}$ given in Eq.~\eqref{eq:HAM}, and $\mathcal{R}$ is the Lellouch-L\"uscher factor, defined as
  \begin{align}\label{eq:LL} 
\mathcal{R}(E_n,\mathbf{P},L)
    \equiv \lim_{E\to E_n}
        \left[
            \frac{E-E_n}{ F(E,\mathbf{P},L)^{-1} + \mathcal{M}(s) }
        \right].
  \end{align}
Applying this identity to the sum of Eq.~\eqref{eq:t_pos} and repeating for $t<0$ one can find,
  \begin{align}\label{eq:ComptonSpec}
&\mathcal{T}_L(p_f,q,p_i,\epsilon)  \\
    &= \sum_{(E_n,\mathbf{q}+\mathbf{p}_f)}
    \frac{e^{i(q^0+\omega_{\mathbf{p}_f}-E_n)T/2-\epsilon T}-1}{q^0+\omega_{\mathbf{p}_f}-E_n+i\epsilon}
    \left\vert \mathcal{R}(E_n,\mathbf{q}+\mathbf{p}_f,L) \right\vert
    \mathcal{A}(s_n,\mathbf{q}^2-(E_n-\omega_{\mathbf{p}_f})^2)
    \left\vert\mathcal{M}(s_n)\right\vert^2
    \mathcal{A}(s_n,\mathbf{q}_{if}^2-(E_n-\omega_{\mathbf{p}_i})^2)
    \nonumber \\
    \ &\ - \sum_{(E_n,\mathbf{p}_i-\mathbf{q})}
    \frac{e^{-i(q^0-\omega_{\mathbf{p}_i}+E_n)T/2-\epsilon T}-1}{q^0-\omega_{\mathbf{p}_i}+E_n-i\epsilon}
    \left\vert \mathcal{R}(E_n,\mathbf{p}_i-\mathbf{q},L) \right\vert
    \mathcal{A}(s_n,\mathbf{q}_{if}^2-(E_n-\omega_{\mathbf{p}_f})^2)
    \left\vert\mathcal{M}(s_n)\right\vert^2
    \mathcal{A}(s_n,\mathbf{q}^2-(E_n-\omega_{\mathbf{p}_i})^2)
    \nonumber, 
  \end{align}
where $\mathbf{q}_{if}=\mathbf{p}_f+\mathbf{q}-\mathbf{p}_i$. 

As was discussed in Sec.~\ref{subsec:review:finvol}, the finite-volume Compton-like amplitudes for $T\to\infty$ are equal to a sum over poles. This spectral decomposition shows that this is true for arbitrary values of $T$. 

In the limit $\epsilon\to0$, the pole in the first sum occurs at $s_n=(q+p_f)^2\equiv s$ while the pole in the second sum occurs at $s_n=(p_i-q)^2\equiv u$. This suggests that in the spectral representation of Eq.~\eqref{eq:ComptonSpec} the first term is closely related to the $s$-channel contribution to the Compton-like scattering process while the second sum corresponds to the $u$-channel contribution. 

\begin{figure}[t]
\centering
\includegraphics[width=0.35\linewidth]{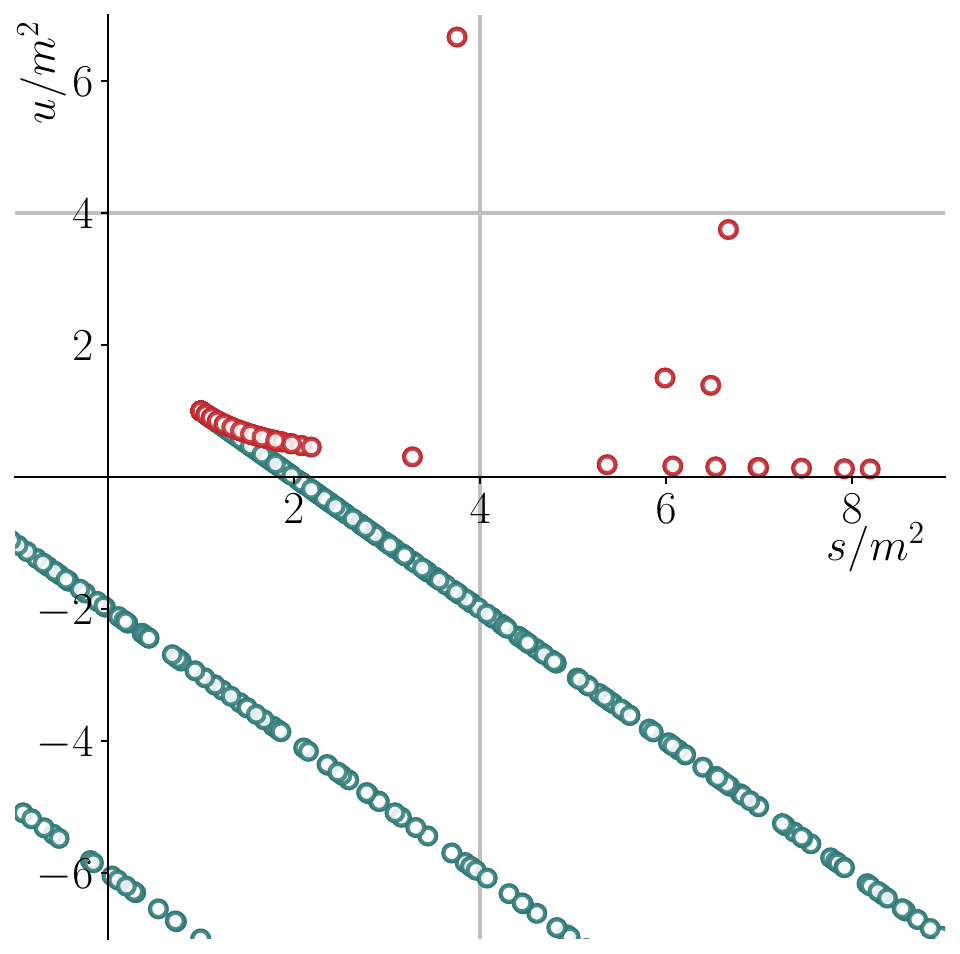}~~~~~~~~~~~~~~~
\includegraphics[width=0.35\linewidth]{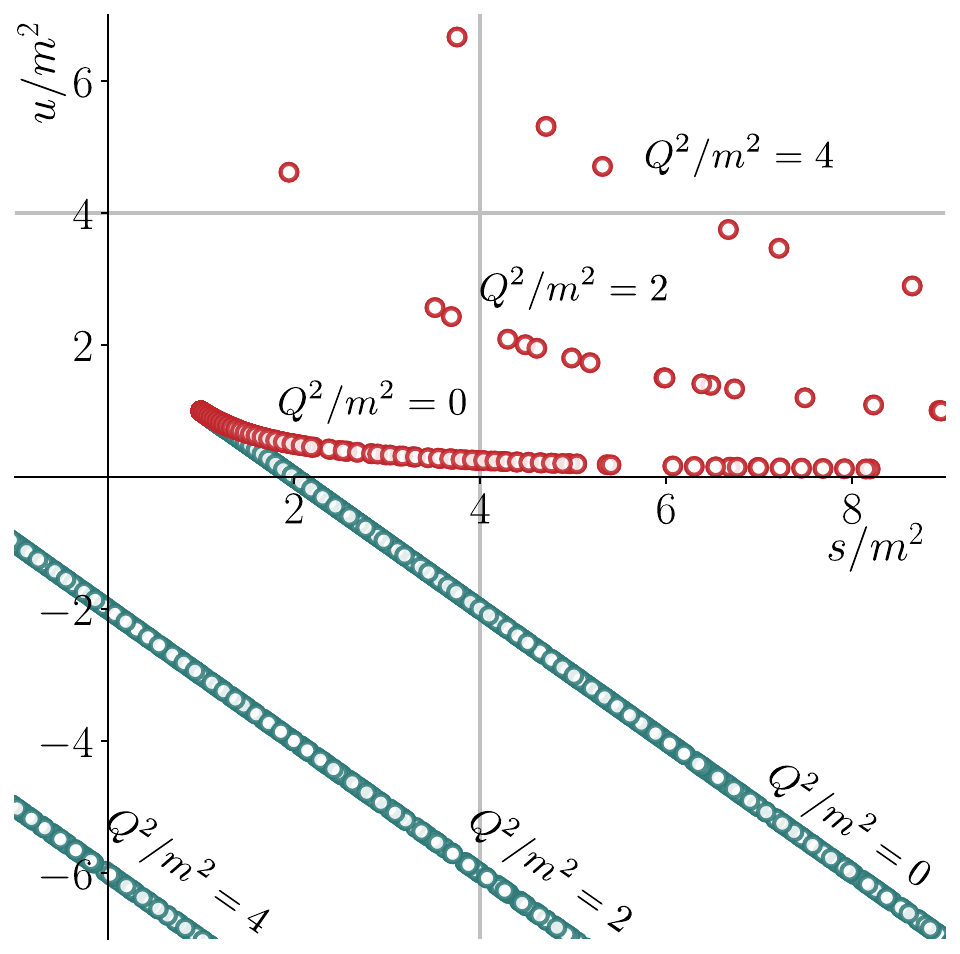} \\[5mm]
\includegraphics[width=0.35\linewidth]{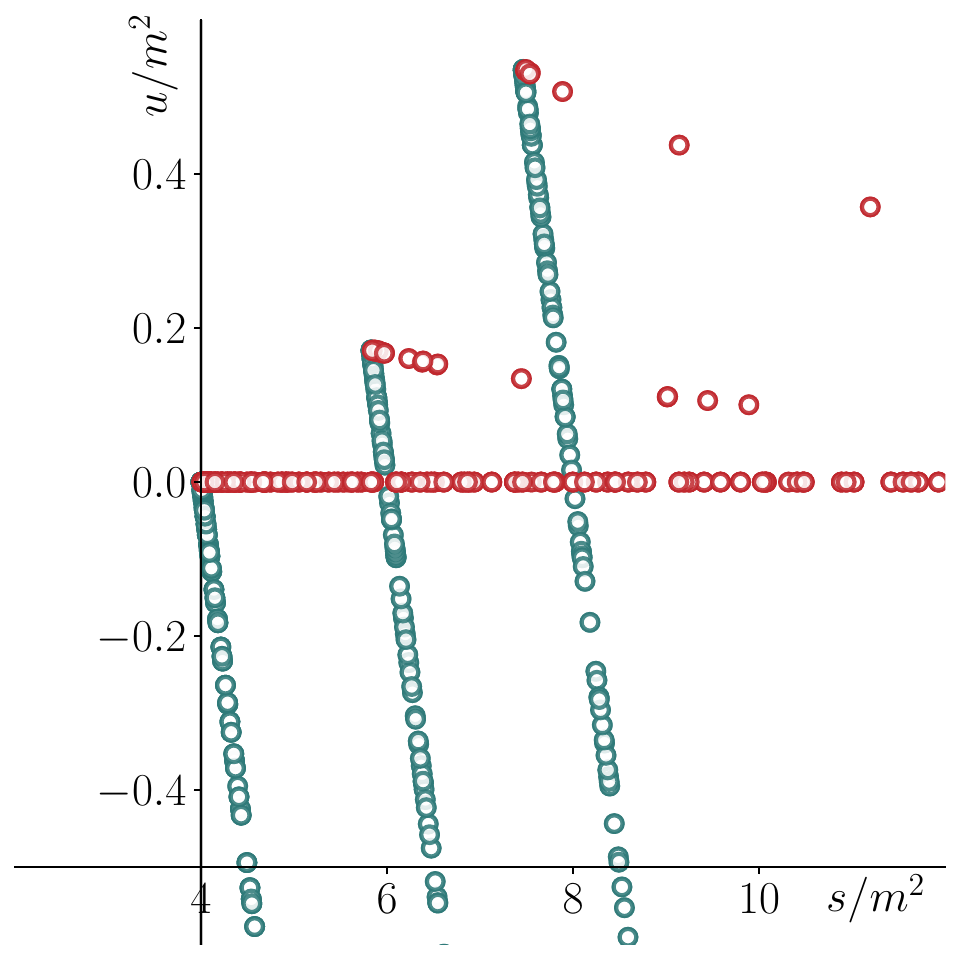}~~~~~~~~~~~~~~~
\includegraphics[width=0.35\linewidth]{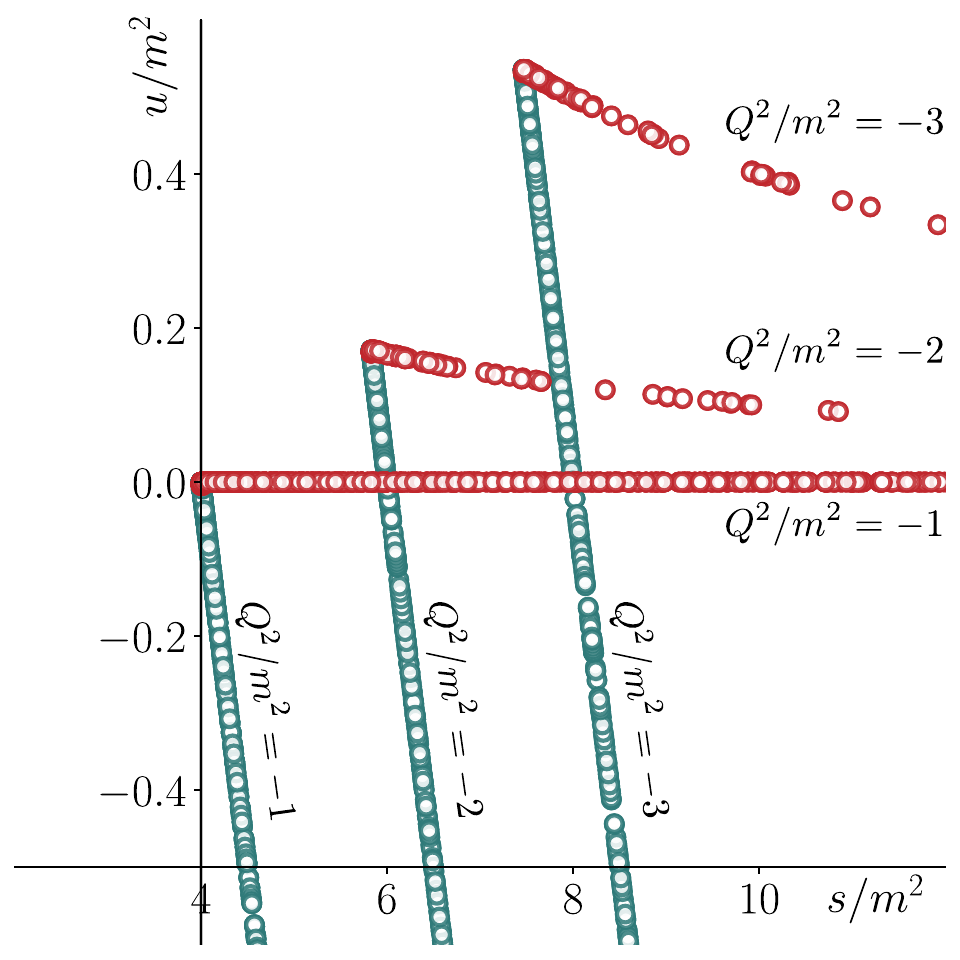}
\caption{The Mandelstam variable $u$ as a function of $s$. The $(s,u_+)$ pairs are shown in red while the $(s,u_-)$ pairs are shown in blue. The left panels show these points for a small volume ($mL=20$) while the right panels consider a larger volume ($mL=40$). The panels on top correspond to kinematics for which $Q^2/m^2=0,2,4$, while the bottom panels cover $Q^2/m^2=-1,-2,-3$. The solid gray lines in the top panels indicate the two-particle threshold for identical particles of mass $m$ for both $s$ and $u$. The value of $Q^2/m^2$ is indicated for the branches of points and we impose $\vert Q^2-Q^2_{if}\vert<0.01m^2$. }  
\label{Fig:Mandelstam_us}
\end{figure}

\subsection{Kinematics in 1+1D}\label{app:kinematics}

Up to this point, we have only considered the $s$-channel contribution of the Compton-like amplitude. Given that the $u$-channel contribution appears in the spectral representation of Eq.~\eqref{eq:ComptonSpec}, we now drop our prior kinematic restrictions to include it in the following analysis. For this, we write the Mandelstam variable $u$ in terms of $s$ and the kinematic variables in the c.m. frame of the initial and final states.  

The final(initial) state can be thought of as a stable particle with mass $m$ and a virtual particle with mass $-Q^2(-Q^2_{if})$. This way, in the c.m. frame, the momenta are
  \begin{align}\label{eq:cmkinematics}
    \begin{array}{rlrl}
p_f^\star&=( \omega_{\mathbf{k}_f}^\star, \mathbf{k}_f^\star ), & 
q^\star&=( \sqrt{s}-\omega_{\mathbf{k}_f}^\star, -\mathbf{k}_f^\star ), \\
p_i^\star&=( \omega_{\mathbf{k}_i}^\star, \mathbf{k}_i^\star ), & 
q_{if}^\star&=( \sqrt{s}-\omega_{\mathbf{k}_i}^\star, -\mathbf{k}_i^\star ).\end{array}
  \end{align}
where $\mathbf{k}_f^\star$ and $\mathbf{k}_i^\star$ are the relative momenta whose magnitudes are given by
  \begin{align}\label{eq:kfmod}
k_f^{\star2} &= \frac{s}{4} + \frac{(m^2+Q^2)^2}{4s} - \frac{m^2-Q^2}{2},
\\ \label{eq:Kimod}
k_i^{\star2} &= \frac{s}{4} + \frac{(m^2+Q^2_{if})^2}{4s} - \frac{m^2-Q^2_{if}}{2}. 
\end{align}
In the limit when $Q^2=Q^2_{if}=-m^2$,  one recovers the result for scattering between two identical particles of mass $m$, $k_f^{\star}=k_i^{\star}=\sqrt{s/4-m^2}$.

The Mandelstam variable $u$ can be obtained from the relation $s+t+u = 2m^2-Q^2-Q^2_{if}$. By evaluating $t$ in the c.m. frame, one gets two possible solutions for $u$, 
  \begin{align}\label{eq:upm}
u_{\pm} 
&= 2\omega_{k_f}^\star\omega_{k_i}^\star \pm 2k_f^\star k_i^\star-s-Q^2-Q^2_{if}, 
  \end{align}
where the positive/negative sign corresponds to the c.m. momenta of the incoming and outgoing particles being anti-parallel/parallel. These expressions simplify a bit further when setting $Q^2_{if}=Q^2$, along with Eqs.~\eqref{eq:kfmod}~and~\eqref{eq:Kimod},
  \begin{align}\label{eq:upmQ2}
u_+=\frac{(m^2+Q^2)^2}{s}, &&
u_-=2m^2-s-2Q^2. 
  \end{align}

By fixing the values of $L$ and $Q^2$, one can determine a set of values for $(s,u_+)$ and $(s,u_-)$. In Fig.~\ref{Fig:Mandelstam_us}, we show two illustrative examples corresponding to two different volumes. In generating this plot, we have fixed $\vert Q^2-Q^2_{if}\vert<0.01m^2$ and the maximal value that we give to $|\mathbf{p}_f|$ and $|\mathbf{p}_i|$ is $2\pi$. The $(s,u_+)$ kinematic points are illustrated as red points, while the $(s,u_-)$ ones are blue. As one can see, there are more blue points points than red ones. In other words, within this setup, one constrains more the region in the Madelstam plane defined by the $(s,u_-)$ curves than the one defined by $(s,u_+)$ lines.

  \begin{figure}
\centering
\includegraphics[width=0.475\linewidth]{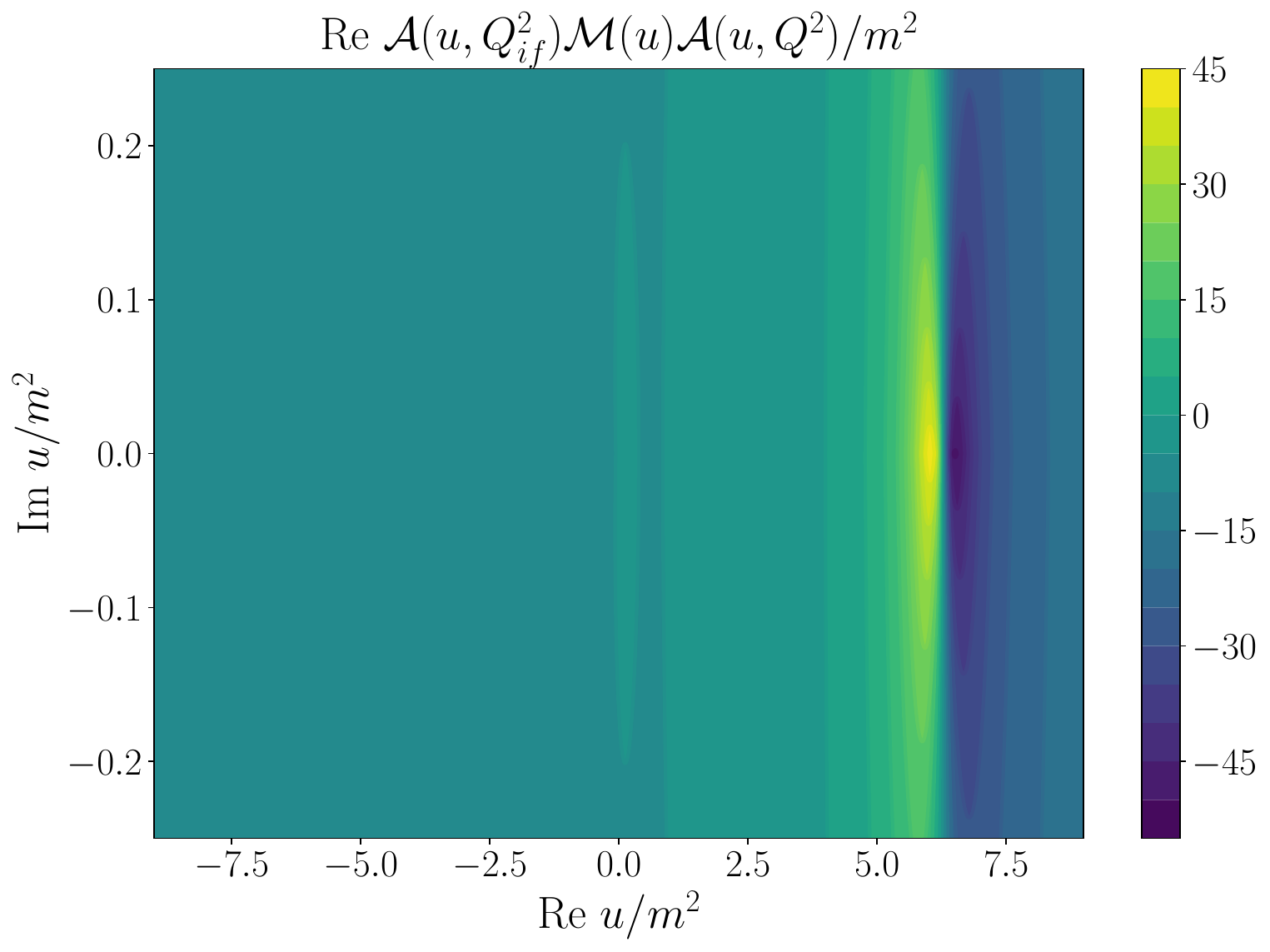}
\includegraphics[width=0.475\linewidth]{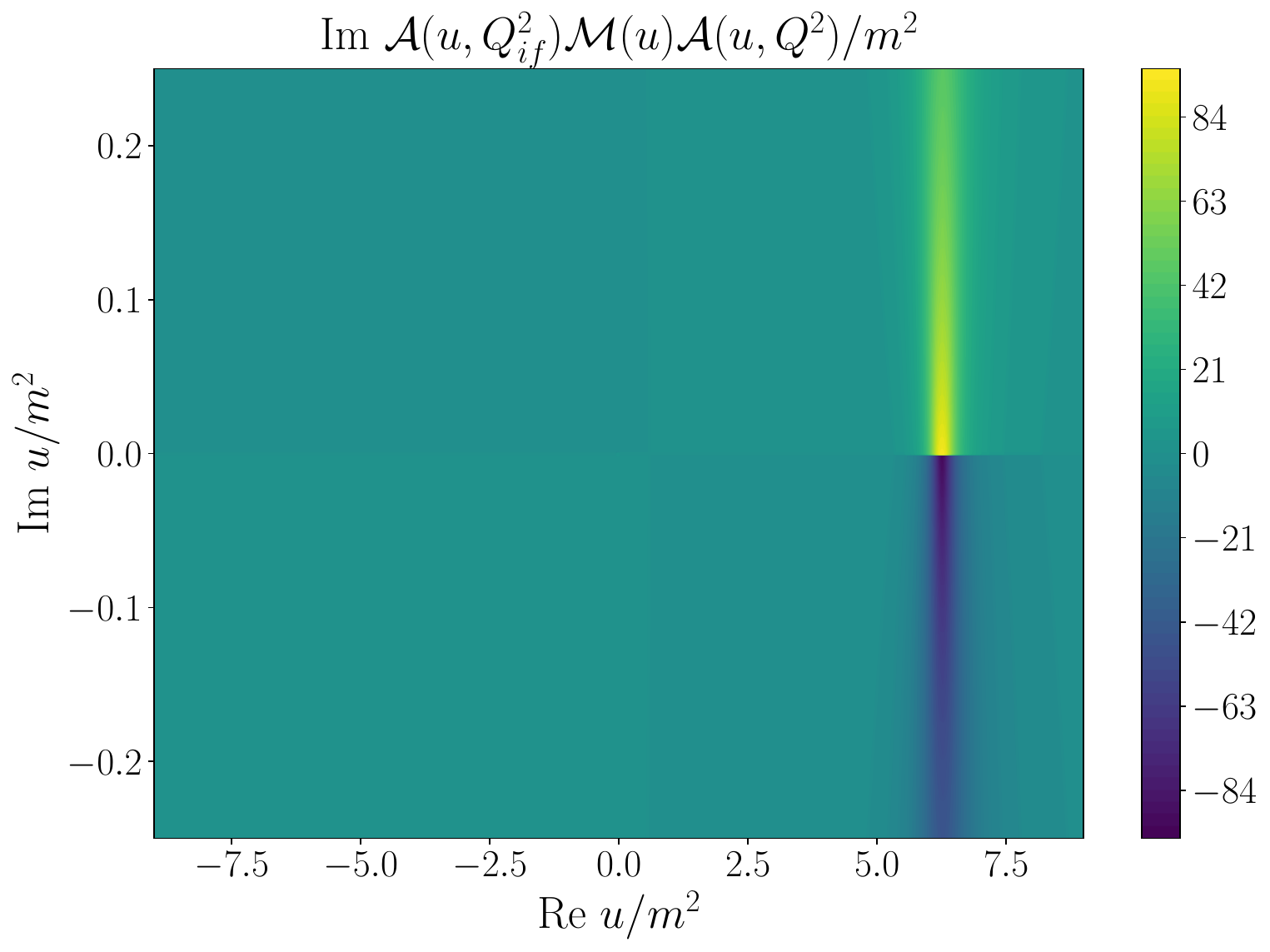}
\caption{Real and imaginary parts of the $u$-channel contribution to the Compton-like amplitude in the complex $u$-plane. The virtualities are fixed to $Q^2=Q^2_{if}=-0.6m^2$. }
\label{fig:T_ucomplex}
  \end{figure}
  
In general, Eqs.~\eqref{eq:upm} and \eqref{eq:upmQ2} are useful for parametrizing the $u$-channel contribution to the Compton-like amplitude. In some limiting cases, the $u_+$ and $u_-$ contributions may be added together. In fact, from Eq.~\eqref{eq:upmQ2} one sees that $u_+=0$ when $Q^2=Q^2_{if}=-m^2$. In this case, the $u_+$ contribution is a constant that can be absorbed into the $w$ term of Eq.~\eqref{eq:ComptonAnalytic}. As a result, one can add the $\mathcal{T}(s,u_+,-m^2,-m^2)$ and $\mathcal{T}(s,u_-,-m^2,-m^2)$ amplitudes trivially. For the estimator $\overline{\mathcal{T}}$, this is equivalent to combining the $u_+$ and $u_-$ kinematics, which improves its statistics. 

Another case of interest is when $Q^2,Q^2_{if}>2m^2$. For these kinematics, the $u_+$-channel contribution can have intermediate on-shell states, as is evident from the $u_+$ definition in Eq.~\eqref{eq:upmQ2}. Because of this potentially singular behavior, it cannot be absorbed into the $w$ term, and it can suffer of larger finite-volume artifacts. And although not immediately evident, because the $u_+$ contribution arises from anti-parallel momenta, it can be harder to estimate statistically using the proposed estimators. By isolating it from the dominant parallel contribution, one can have a better systematic control of the $u_+$ piece. 

From Fig.~\ref{Fig:Mandelstam_us}, one can also notice that $u=u_-$ goes rapidly into the $u_-<0$ region where a left-hand cut is present. This is due to a $\sqrt{u}$ in the phase-space function in Eq.~\eqref{eq:rho}. Given that the scattering amplitudes have been projected on-shell over the two-particle threshold, thus over the right-hand cut region, these do not have an analytic handle outside of it. In other words, the evaluation of these amplitudes along the left-hand cut region, i.e. in  $u=u_-$, can be absorbed into the $w$ term of Eq.~\eqref{eq:TwAMA}.

As an illustration of the singular structure of the $u$-channel contribution to $\mathcal{T}$ in the complex $u$-plane, given by $\mathcal{A}(u,Q^2_{if})\mathcal{M}(u)\mathcal{A}(u,Q^2)$, in Fig.~\ref{fig:T_ucomplex} we show this term for an specific model. In particular, we use the same resonant K-matrix form for $\mathcal{M}$, and we a simple pole in $Q^2$ for $\mathcal{A}$. For this example, the virtualities are fixed to $Q^2=Q^2_{if}=-0.6m^2$. One can see that for these models, the only singularities in the $u$-channel are due to the right-hand branch cut and the nearby resonant pole in ${\rm Re}~u\sim m_R^2$, which are only accessible for $Q^2,Q^2_{if}>2m^2$, as can be seen from Eq.~\eqref{eq:upmQ2} and Fig.~\ref{Fig:Mandelstam_us}. Although, we show this for a choice of $Q^2$, with $Q^2_{if}=Q^2$, the same behavior is observed regardless of these values. 

\subsection{Estimators in a finite spacetime}\label{subsec:FinTEstimator}

\begin{figure}[t]
    \centering
    \includegraphics[width=0.95\linewidth]{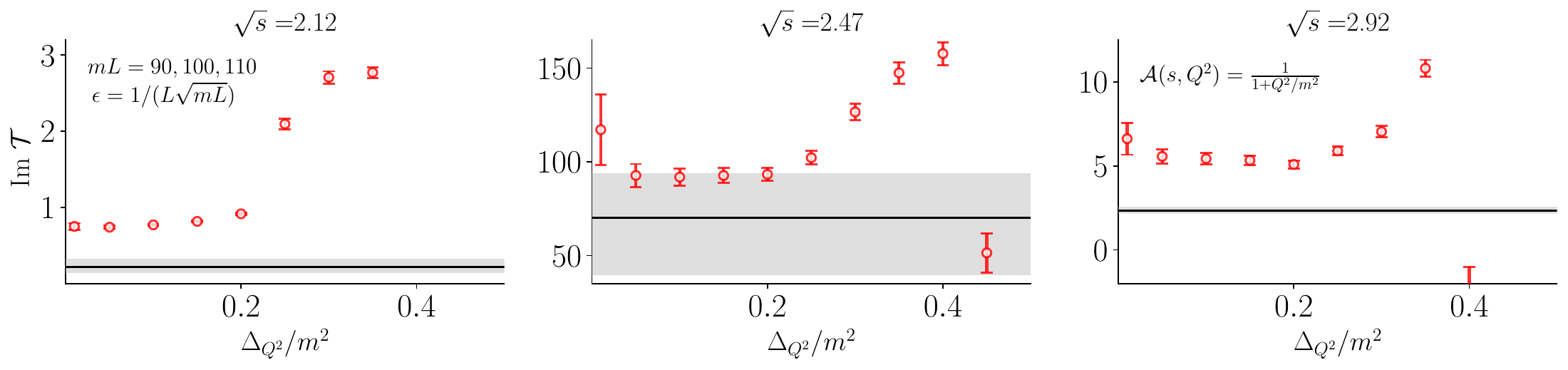}
    \caption{Imaginary part of the Compton-like amplitude for various c.m. energies. The red points represent the estimators, the solid line shows the infinite-volume value, and the band shows the maximum and minimum values of ${\rm Im}~\mathcal{T}$ for a wavepacket with mean energy $\sqrt{s}$. For this example the mean values of the virtualities are $Q^2_{if}=Q^2=-0.6m^2$. }
    \label{fig:DQ2_convergence}
\end{figure}

Since the Mandelstam variable $u$ is a multi-valued function of $s$, the set of kinematics per volume is divided into the two subsets discussed in the previous section,
  \begin{align}\label{eq:setspm}
\left\{\mathbf{p}_f,\mathbf{q},\mathbf{p}_i\right\}_{+}&=
\left\{\mathbf{p}_f,\mathbf{q},\mathbf{p}_i
    ~|~\mathbf{k}^\star_f\cdot\mathbf{k}^\star_i<0\right\}, &
\left\{\mathbf{p}_f,\mathbf{q},\mathbf{p}_i\right\}_{-}&=
\left\{\mathbf{p}_f,\mathbf{q},\mathbf{p}_i
    ~|~\mathbf{k}^\star_f\cdot\mathbf{k}^\star_i\geq0\right\},
  \end{align}
where $\mathbf{k}^\star$ represents the c.m. frame linear momentum. In consequence, we have to consider two cases for the estimator of the Compton-like amplitude
  \begin{align}\label{eq:estTFT}
\overline{\mathcal{T}}(s,u_{\pm},Q^2,Q^2_{if})
    = \frac{1}{\mathcal{N}} \sum_L\sum_{\{\mathbf{p}_f,\mathbf{q},\mathbf{p}_i\}_{\pm}}
      \mathcal{T}_L(p_f,q,p_i,\epsilon).
  \end{align}
Using Eq.~\eqref{eq:ComptonSpec} in Eq.~\eqref{eq:estTFT}, the estimator can now be tested under finite-$T$ conditions. 

The parametrization of the K matrix in this section is set by $g=2.5$, $m_R=2.5m$, and $h^{(0)}=0$. We consider two models for the parametrization of $\mathcal{A}$, one where it is a smooth function of $Q^2$ and one where it has a single particle pole at $Q^2=-m^2$. For simplicity, we keep $Q^2_{if}=Q^2$. The $\epsilon$-prescription is $\epsilon=1/(L\sqrt{mL})$ for all cases. 

The parameters for the boost averaging are given by $\Delta_{Q^2}=0.25m^2$ and $\Delta_{\sqrt{s}}=0.05m$. The choice of $\Delta_{Q^2}$ here is motivated by the behavior of the estimator as a function of this parameter. Mainly, when $\mathcal{A}$ is a singular function of $Q^2$, close to the pole the estimators are increasingly inconsistent with the infinite-volume amplitude at $\Delta_{Q^2}\gtrsim0.25m^2$. This behavior is shown for the estimators at different c.m. energies in Fig.~\ref{fig:DQ2_convergence}. When $\mathcal{A}$ is a smooth function of $Q^2$, the estimator does not change significantly with $\Delta_{Q^2}$. 

Since we are comparing different representations of the Compton-like amplitude, see Eqs.~\eqref{eq:TwAMA} and~\eqref{eq:ComptonSpec}, we only show their imaginary parts. By unitarity, both are assured to be equal thus allowing for a comparison between the estimators under finite-$T$ conditions and the infinite-volume limit. 

\begin{figure}[t]
    \centering
    \includegraphics[width=0.95\linewidth]{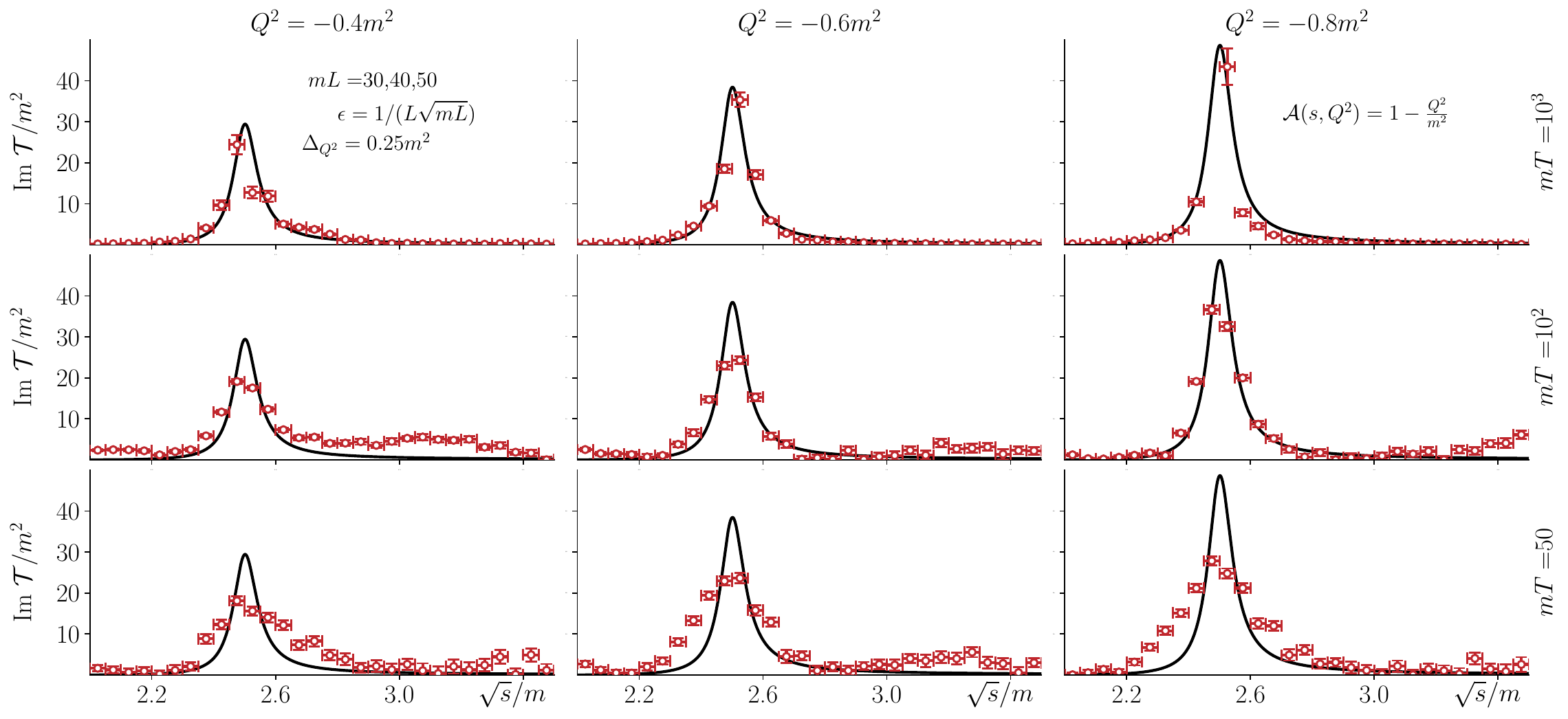}
    \caption{Imaginary part of the Compton-like amplitude. From top to bottom, the temporal extent of the spacetime is decreased from $mT = 10^3$ to $50$ as indicated on the right side. The limit $Q^2\to-m^2$ is approached from left to right, and $Q^2_{if}=Q^2$. The volumes employed in all cases are indicated in the top-left panel. The red dots show the estimators and the line shows the infinite-volume amplitude. Since $Q^2\sim-m^2$, the subsets of kinematics in Eq.~\eqref{eq:setspm} are combined into a single set.}
    \label{fig:Spectral_MultimL_smooth}
\end{figure}

\subsubsection{Model without single-particle singularities}

In this case the parametrization of $\mathcal{A}$ is given by $h^{(1)}=0$, $h^{(2)}=1$, and $h^{(3)}=1-Q^2/m^2$. Figure~\ref{fig:Spectral_MultimL_smooth} shows our results for the estimator $\overline{\mathcal{T}}$ using different values of $mT$ and $Q^2\sim-m^2$. The red points correspond to the estimators using the spectral representation in Eq.~\eqref{eq:ComptonSpec}. The black line corresponds to the infinite-volume on-shell representation in Eq.~\eqref{eq:ComptonAnalytic}. 

\begin{figure}[t]
    \centering
    \includegraphics[width=0.95\linewidth]{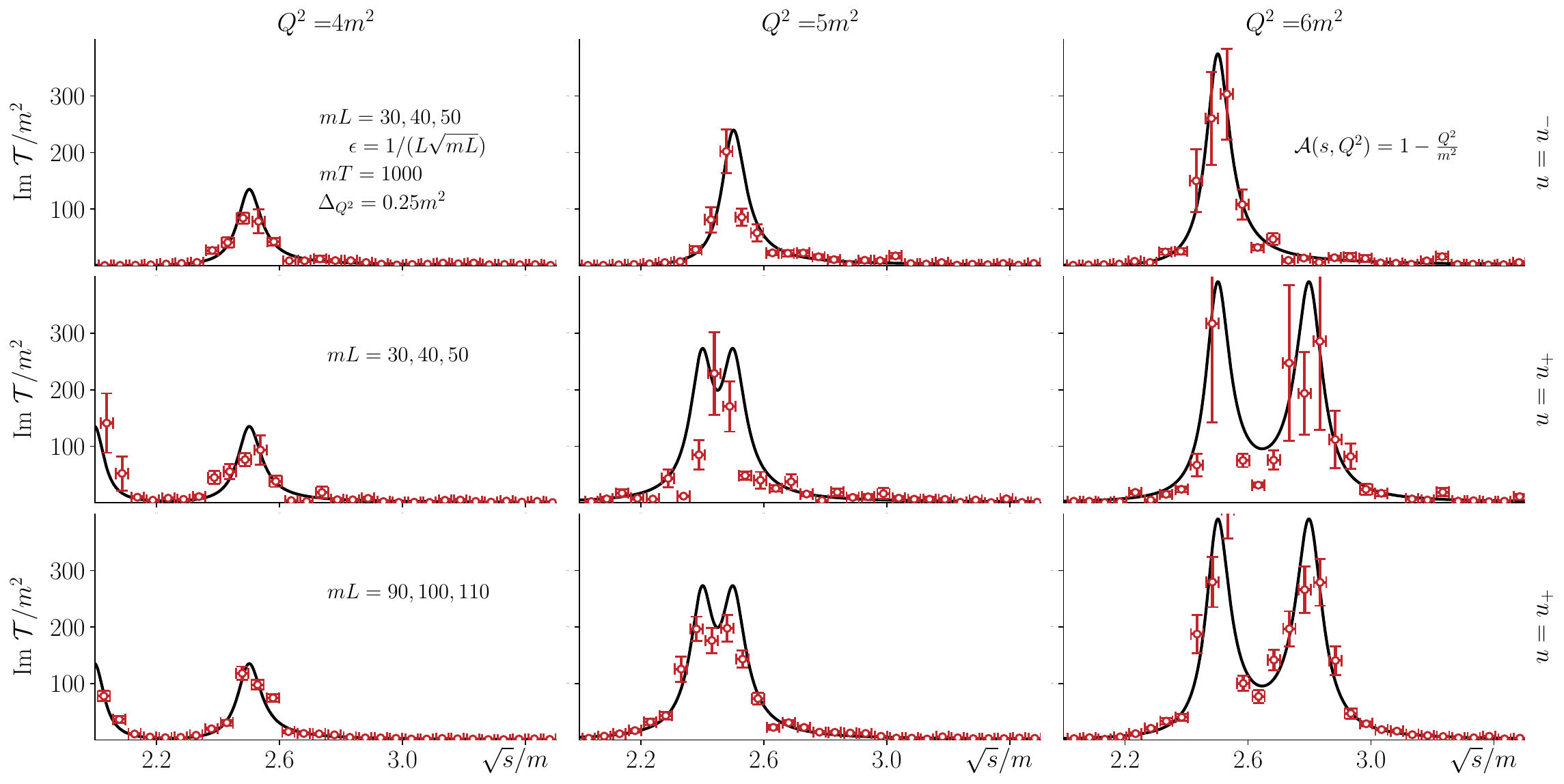}
    \caption{Imaginary part of the Compton-like amplitude. Same symbols as in Fig.~\ref{fig:Spectral_MultimL_smooth} but for large, positive values of $Q^2$, also $Q^2_{if}=Q^2$. Since the virtualities are such that $Q^2>-2m^2$, the kinematics described by $u=u_+$ and $u=u_-$ are treated separately, as indicated in each row of panels. The time extension considered in these examples is indicated in the top-left panel.}
    \label{fig:Spectral_medQ2_smooth}
\end{figure}

There are two key observations to be made from these results. First, as the temporal extent increases, the estimator recovers the desired amplitude more accurately. This suggests that one can systematically determine errors associated with the truncation of $T$ by considering larger temporal extents. Secondly, for smaller time extents and close to the singularities, the estimator fails to reproduce the amplitude. This is most evident near the peaks of the amplitudes in Figs.~\ref{fig:Spectral_MultimL_smooth} and \ref{fig:Spectral_medQ2_smooth}, which are dynamically generated by pole singularities. This is the same behavior observed for finite-volume artifacts. Although we only illustrate this for one representative example, we find the same conclusion for all models with smooth parametrizations of $\mathcal{A}$.

Another case of interest is when both $s$ and $u=u_+$ are above the two-particle threshold, which occurs at $Q^2>2m^2$. In this case, the kinematics of $u=u_+$ and $u=u_-$ have to be treated separately. Figure~\ref{fig:Spectral_medQ2_smooth} shows the estimator $\overline{\mathcal{T}}$ under these conditions. It is observed that the $u=u_+$ contribution manifests as a second peak in the amplitude. To observe this behavior, it is necessary to increase the order of magnitude of the physical volumes $mL$, while maintaining the same temporal extent $mT$. 

\subsubsection{Model with a single-particle singularity}

\begin{figure}[t]
    \centering
    \includegraphics[width=0.95\linewidth]{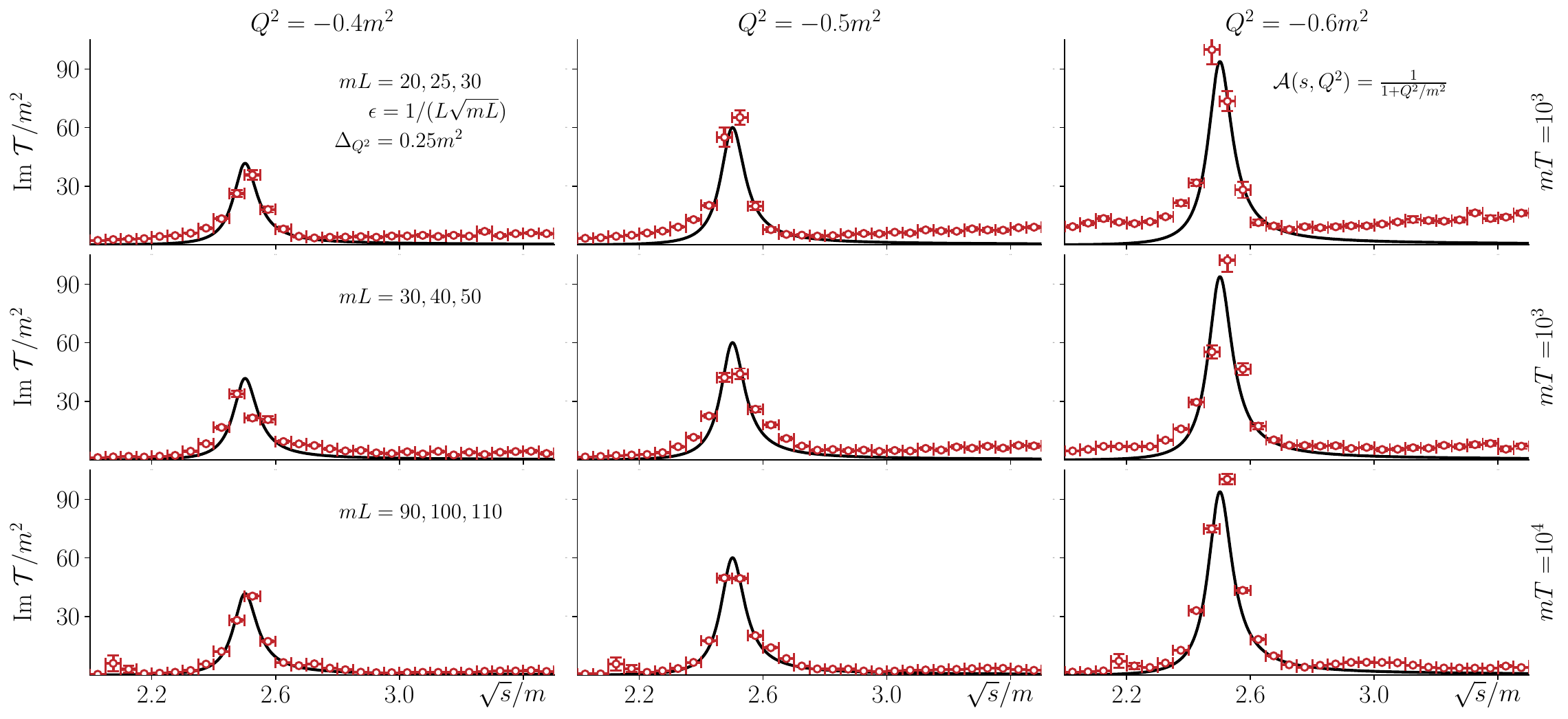}
    \caption{Imaginary part of the Compton-like amplitude. Same symbols as in Fig.~\ref{fig:Spectral_MultimL_smooth} but for a model that admits a single-particle pole at mass $m$, we keep $Q^2_{if}=Q^2$. From left to right the virtuality approaches the pole. The volumes $mL$ used in each row are indicated in the left-most panels and the time extensions $mT$ are indicated on the right side. }
    \label{fig:Spectral_singular}
\end{figure}

In this case the parametrization of $\mathcal{A}$ has a pole corresponding to a particle of mass $m$, this is set by making $h^{(1)}=m^2$, $h^{(2)}=1$, and $h^{(3)}=0$. Figure~\ref{fig:Spectral_singular} shows the estimator $\overline{\mathcal{T}}$ for virtualities approaching the pole, $Q^2\to-m^2$, and different combinations of $mL$ and $mT$. In contrast to previous results, there is a gap between the estimator and the infinite volume amplitude far from the resonant energy at $\sqrt{s}=m_R$. As the virtuality approaches the single-particle pole, this gap decreases more slowly with the physical volume $mL$. This behavior in the estimator can be understood from the sum over the two-particle spectrum in Eq.~\eqref{eq:ComptonSpec} which requires evaluating $\mathcal{A}$ at points close to its singular values. Our observations indicate that, to faithfully recover the amplitude, one may need to consider larger volumes and time extensions as the single-particle pole is approached. 

\section{Conclusion and outlook}\label{sec:conclusion}

We have continued the numerical investigation of estimators for physical scattering amplitudes based on correlation functions set in a finite Minkowski spacetime. These estimators were first introduced in Ref.~\cite{Briceno:2020rar} for Compton-like scattering processes. In particular, we have addressed three new features. First, we have tested the conjecture, made first in Ref.~\cite{Briceno:2020rar}, that the LSZ reduction formula can be used to construct estimators for transition and purely hadronic scattering amplitudes. Secondly, we tested the estimators for systems with multiple coupled channels. Finally, we have explored the effects that finite time extensions have on the estimator $\overline{\mathcal{T}}$. 

The results shown here suggest the applicability of these estimators to energy regions where many two-particle states may go on shell. We also constrain the order of physical volumes and time separations necessary to recover amplitudes in 1+1D within reasonable error. For the most part, this is achieved for spatial volumes of order $mL\sim\mathcal{O}(10-10^2)$, under either infinite or finite time separations. 

For models where the vertex function $\mathcal{A}$ of transition amplitudes is a smooth function of $Q^2$, physical time separations of order $mT\sim\mathcal{O}(10^2-10^3)$ suffice to have an agreement between estimators and amplitudes, within errors. Still, for models where the current can go on-shell into a particle of mass $m$, time extents as big as $mT\sim\mathcal{O}(10^4)$ may be necessary for resolving the scattering amplitudes. The increment in the order of $mT$ needed for the latter case is due to the singular behavior of $\mathcal{A}$ as one approaches its single-particle pole. 

It is worth emphasizing that the procedure outlined here provides a straightforward method for estimating the statistical error. Although assessing systematic errors generally requires repeating the calculations for different spacetimes, the procedure is systematically improvable. In other words, the systematic error on the estimators will decrease with increasing spacetime volumes.

Although we have shown that the purely hadronic amplitudes can indeed be obtained following the outlined procedure, the results are dependent on how close one chooses to approach the single-particle poles in $\mathcal{A}$. Ultimately, the amplitudes can not depend on this. A possible systematic procedure to resolve this unaccounted  error is the use of Cauchy's theorem. At this point, it is unclear how that may be implemented in the current setup.

Our study focused on estimators and amplitudes in 1+1D, the case for 3+1D can be implemented using the same expressions presented in Sec.~\ref{sec:review}  after replacing the $\rho$ and $F$ functions with their 3+1D analogues. Similarly, one can adopt the techniques being applied to study three-particle systems, to further test the conjecture that these estimators should work independent of the kinematics and/or the number of intermediate particles that can go on shell. 

Further studies are necessary to find optimal values of some parameters  ($\epsilon,~Q^2,~\Delta_{Q^2},~\Delta_{\sqrt{s}}$) that could improve the correspondence between estimators and amplitudes. For example, for $mL$ given, the $\epsilon$-prescription has to be large enough to smoothen the real singularities in the finite-volume amplitudes and small enough so it does not suppress the estimator. Likewise, as the estimator struggles to recover the amplitude near singular points (e.g. poles or threshold singularities), one could opt for a smaller value of $\Delta_{\sqrt{s}}$ around these regions. However, since this condition implies averaging over less kinematics as a result, the finite-volume effects would be less suppressed. 

Finally, although we have presented further empirical evidence supporting the fact that the finite-volume estimators have reasonable constraints of infinite-volume amplitudes in arbitrary kinematics, a general proof of the correspondence between estimators and amplitudes is still missing.

\section*{Acknowledgements}

MAC and RAB acknowledge the support of the USDOE Early Career award, contract DE-SC0019229. MAC is supported by the ODU PhD Physics program and the 2023-2024 JSA/JLab Graduate Fellowship Program. MAC was also supported by the Jefferson Lab LDRD project LD2117. RAB was supported in part by the U.S. Department of Energy, Office of Science, Office of Nuclear
Physics under Awards No. DE-AC02-05CH11231. AMS is supported in part by the DOE Office of Science, under ECA Grant No. DE-SC0023047, and in part by the Virginia Space Grant Consortium, under Grant No. 23-157-100846-010/80NSSC20M0056. We thank J. V. Guerrero, A. Rodas, A. W. Jackura, F. Ringer, and Z. Davoudi for the useful discussions. 

\nocite{*}
\bibliographystyle{apsrev4-1}
\bibliography{References}

\end{document}